\documentclass[final,aps,pra,10pt,nofootinbib,longbibliography,showkeys]{revtex4-2}

\usepackage{graphicx}
\usepackage{amsmath}
\usepackage{amssymb}
\usepackage{microtype}
\usepackage{booktabs}
\usepackage{sidecap}
\usepackage{enumitem}

\usepackage{tikz}    
\usepackage{orcidlink}

\usepackage{hyperref}

\definecolor{myblue}{HTML}{2222BB}

\hypersetup{
colorlinks=true,
allcolors=myblue,
pdfborder={0 0 0}
}

\usepackage{bm}  

\renewcommand{\Re}[1]{\mathop{\mathrm{Re}}#1}
\renewcommand{\Im}[1]{\mathop{\mathrm{Im}}#1}

\begin{document}

\title{Semiconductor Laser Linewidth Theory Revisited}

\author{Hans Wenzel\,\orcidlink{0000-0003-1726-0223}}
 \email{hans.wenzel@fbh-berlin.de}
 \affiliation{Ferdinand--Braun--Institut gGmbH, Leibniz--Institut f\"ur 
 H\"ochstfrequenztechnik, Gustav--Kirchhoff--Str. 4, 12489 Berlin, Germany
}

\author{Markus Kantner\,\orcidlink{0000-0003-4576-3135}}%
\author{Mindaugas Radziunas\,\orcidlink{0000-0003-0306-1266}}%
\author{Uwe Bandelow\,\orcidlink{0000-0003-3677-2347}}%
\affiliation{%
 Weierstrass Institute for Applied Analysis and Stochastics\\
 Mohrenstr. 39, 10117 Berlin, Germany
}

\begin{abstract}
More and more applications require semiconductor lasers distinguished not only by large modulation bandwidths or high output powers, but also by small spectral linewidths. 
The theoretical understanding of the root causes limiting the linewidth is therefore of great practical relevance. 
In this paper, we derive a general expression for the calculation of the spectral linewidth step by step in a self--contained manner.
We build on the linewidth theory developed in the 1980s and 1990s but look from a modern perspective, in the sense that we choose as our starting points the time-dependent coupled--wave equations for the forward and backward propagating fields and an expansion of the fields in terms of the stationary longitudinal modes of the open cavity.
As a result, we obtain rather general expressions for the longitudinal excess factor of spontaneous emission ($K$--factor) and the effective $\alpha$--factor including the effects of nonlinear gain (gain compression) and refractive index (Kerr effect), gain dispersion, and longitudinal spatial hole burning in multi--section cavity structures.
The effect of linewidth narrowing due to feedback from an external cavity often described by the so--called chirp reduction factor is also automatically included.  
We propose a new analytical formula for the dependence of the spontaneous emission on the carrier density avoiding the use of the population inversion factor.
The presented theoretical framework is applied to a numerical study of a two--section distributed Bragg reflector laser.
\end{abstract}


\maketitle


\section{Introduction}

Many applications of semiconductor lasers utilized in miniaturized contemporary photonic integrated devices for coherent optical communication, optical atomic clocks, atom interferometry, gravitational wave detection, space--based metrology, and optical quantum sensing impose strict requirements on the coherence of the light source, which can be expressed in terms of the spectral linewidth. The theoretical understanding of the governing factors limiting the linewidth is therefore of great practical relevance. After a first prediction by Schawlow and Townes \cite{schawlow1958infrared}, the corresponding theoretical framework was developed in the 1960s \cite{haken1964nonlinear,lax1967classical,haug1967theory,scully1967quantum,risken1968statistik}). 
Later, in the 1980s and 1990s with the rapidly advancing technology of the fabrication of semiconductor lasers and their applications the theory of the spectral linewidth was further refined.
The following 5 milestones can be distinguished:

\begin{enumerate}[itemsep=1pt,topsep=1pt,label=(\roman*)]
\item The first milestone is the discovery of the enhancement of the fraction of the spontaneous emission going into the lasing mode in gain--guided lasers and the derivation of a corresponding excess factor ($K$--factor) by Petermann \cite{petermann1979calculated}.
Siegman recognized this effect as a general property of non--Hermitian laser
cavities \cite{siegman1989excess}.
Later his discussion of the power--nonorthogonality of the transversal modes and its consequences was extended to the case of the power--nonorthogonality of the longitudinal modes of laser cavities \cite{hamel1989nonorthogonality}. 
In Ref. \cite{wenzel1996mechanisms} it was discovered, that the longitudinal modes can become degenerate for certain parameter configurations resulting in an infinite $K$--factor.  
The occurrence of such \emph{exceptional points} is not restricted to lasers but is inherent to non--Hermitian systems. See \cite{ozdemir2019parity} for a recent review.

\item The second milestone is the discovery of a linewidth enhancement in semiconductor gain materials 
caused by refractive index fluctuations in response to fluctuations of the carrier density. 
Due to gain clamping, intensity fluctuations (which have negligible direct effect on the linewidth) can cause substantial refractive index changes, which in turn lead to fluctuations of the phase. The magnitude of this \emph{amplitude-phase coupling} is quantified by the linewidth enhancement factor $\alpha_\mathrm{H}$ or $\alpha$--factor introduced by Henry~\cite{henry1982theory}. Later it was found that in DFB lasers the $\alpha$--factor has to be replaced by an effective factor $\alpha_\mathrm{H,eff}$ \cite{amann1990linewidth} and a general expression for $\alpha_\mathrm{H,eff}$ valid for distributed feedback (DFB), distributed Bragg reflector (DBR) and external cavity lasers was derived in \cite{tromborg1991theory}.

\item The third milestone is the discovery of the possibility of linewidth reduction due to optical feedback from an external cavity
by Patzak \textit{et al.} \cite{patzak1983semiconductor}. Later a relation to the reduction of the frequency chirp was established 
\cite{kazarinov1987relation, tromborg1987transmission}.

\item The forth milestone is related to the deterioration of the linewidth due to charge carriers injected into a phase tuning section within the cavity.  Amann and Schimpe figured out that carrier noise is the origin resulting in an additional contribution to the linewidth, if the $\alpha$--factors are different in the gain and phase tuning sections \cite{amann1990excess}.

\item The last milestone is the discovery of the  enhancement of the linewidth due to fluctuations of the shape of the profile of the optical power in the cavity by Tromborg and co-workers \cite{tromborg1991theory}, which is particularly important in the vicinity of instabilities \cite{olesen1992mode, schatz1992longitudinal}.
The most sophisticated linewidth theory including fluctuations of the shape of the power profile was published in \cite{olesen1993stability, tromborg1994traveling}. 
\end{enumerate}

In this paper, we will derive a semi-analytical expression of the spectral linewidth from a modern perspective step by step in a self--contained manner collecting all necessary ingredients that are otherwise found only scattered in the literature.
The underlying theoretical approach is the classical Langevin formalism \cite{lax1966classical} where the deterministic equations for the optical field and the carrier density are supplemented by noise sources (Langevin forces). By means of these stochastic terms, quantum field theoretical phenomena (in particular spontaneous emission), which are essential for the laser linewidth, can be adequately treated within the framework of the semi-classical theory. The correlation functions of the Langevin noise sources are determined by the fluctuation-dissipation theorem, for which we refer to Refs.~\cite{lax1966quantum, marani1995spontaneous, henry1996quantum}.

While the starting point of most authors is the Helmholtz equation and an expansion of its Green's function \cite{tromborg1991theory, henry1986theory}, it is more transparent to start from the time-dependent coupled--wave equations and to expand the forward and backward propagating fields in terms of the stationary longitudinal modes of the open cavity \cite{wenzel1996mechanisms} yielding simpler expressions.
The linewidth expression derived in such a manner includes automatically the findings described as milestones (i)--(iii), \textit{i.e.}, longitudinal $K$--factor, effective $\alpha$--factor, and chirp reduction factor.
Additionally, we propose a new analytical formula for the dependence of the spontaneous emission on the carrier density, which is independent of the commonly employed \emph{population inversion factor}, \textit{i.e.}, the ratio between the spontaneous rate of downward band--to--band transitions and the stimulated rate of downward and upward transitions \cite{lasher1964spontaneous}. 
The new formula avoids problems with the singularity of the population inversion factor near the transparency density and is in good agreement with microscopic calculations, see Sec.~\ref{sec:spont}.
The theory can be easily extended to include carrier noise (iv), too, as sketched in the Outlook in Sec.~\ref{sec:outlook}.

\section{Prerequisites and basic assumptions}
\label{sec:2}

\begin{SCfigure}[][t]
\includegraphics[scale=1.0]{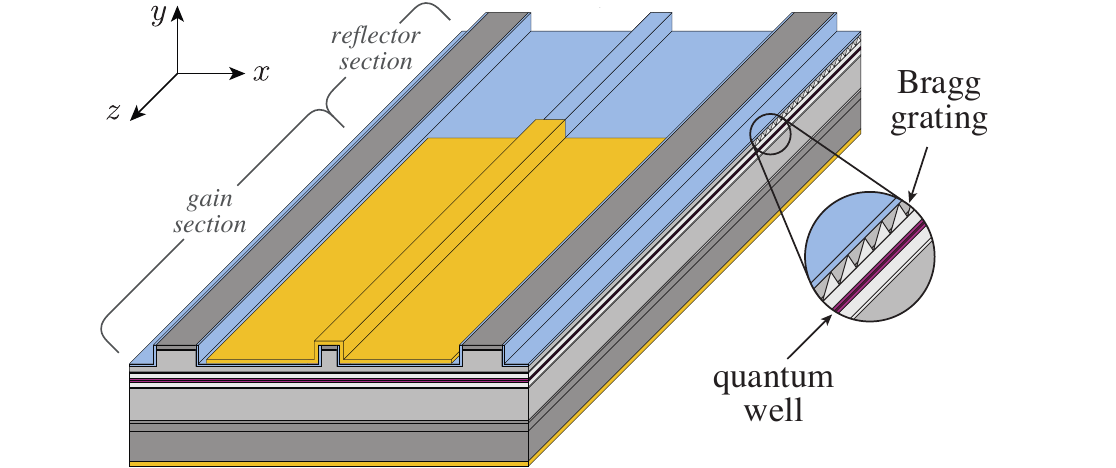}
\caption{Schematic view of a two--section DBR laser consisting of an electrically biased gain section and a passive Bragg reflector section, exemplifying an edge--emitting multi--section semiconductor laser.\vspace{0ex}
\label{fig:sketch}}
\end{SCfigure}  

We consider edge--emitting semiconductor lasers as sketched in Fig. \ref{fig:sketch} consisting of an arbitrary number of sections of different functionality. 
The transverse cross section is uniform within each section, but may vary from section to section.
The preferred propagation direction of the optical field under lasing conditions is along the cavity axis parallel to the longitudinal coordinate $z$. 
Thus laser can be well described by the traveling--wave ansatz for the main component of the electric field strength
\begin{equation}
\label{eq:TWAnsatz}
E(x,y,z,t)=\frac{e(x,y)}{\sqrt{2\varepsilon_0cn}}\left[\Psi^+(z,t)\mathrm{e}^{-i\beta_0z}+\Psi^-(z,t)\mathrm{e}^{i\beta_0z}\right]\mathrm{e}^{i\omega_0t}+\text{c.c.},
\end{equation}
employing a scalar approximation and neglecting the dynamics in the transverse $(x,y)$--plane.
This means that we assume \emph{index-guiding} in the transverse (lateral--vertical) plane and lasing in a single transverse mode and exclude \textit{gain--guiding}, as appropriate for narrow--linewidth lasers. The real--valued field distribution $e(x,y)$ is obtained as a solution of a real--valued waveguide equation normalized to $\iint e^2(x,y)\,dxdy=1$ \cite{agrawal1993}.
Moreover, $\varepsilon_0$ is the vacuum permittivity, $c$ the vacuum speed of light, $n$ the reference modal index, $\beta_0=2\pi n/\lambda_0$ the reference propagation constant, $\lambda_0$ the reference wavelength and $\omega_0=2\pi c/\lambda_0$ the reference frequency. The prefactor is chosen such that $\Vert\Psi\Vert^2 \equiv |\Psi^+|^2+|\Psi^-|^2$ is the optical power $P$ with unit W.
The left and right propagating fields $\Psi^\pm$ vary slowly both in the longitudinal coordinate $z$ and time $t$ and fulfill the stochastic time--dependent coupled--wave equations \cite{radziunas2017traveling, sieber1998travelling}
\begin{equation}
\label{eq:cwe}
\frac{n_{\mathrm{g}}(z)}{c}
\frac{\partial}{\partial t}\Psi(z,t)
+\left[\sigma_z \frac{\partial}{\partial z} + iM(z,N,\Vert\Psi\Vert^2) + i\widehat{\cal D}(z,t)\right]\Psi(z,t)
=F_\mathrm{sp}(z,t,N)
\end{equation}
with
\begin{equation*}
\Psi(z,t) =
\begin{bmatrix}
\Psi^+(z,t) \\
\Psi^-(z,t)
\end{bmatrix},\qquad
\sigma_z=\begin{bmatrix}
1  &
0 \\
0 &
-1
\end{bmatrix},
\end{equation*}
and
\begin{subequations}
\begin{align}
M(z,N,\Vert\Psi\Vert^2)&=\begin{bmatrix}
\Delta\beta(z,N,\Vert\Psi\Vert^2)  &
\kappa^{+}(z) \\
\kappa^{-}(z) &
\Delta\beta(z,N,\Vert\Psi\Vert^2)
\end{bmatrix},\\
\Delta\beta(z,N,\Vert\Psi\Vert^2) &=\beta(z,N,\Vert\Psi\Vert^2)-\beta_0
=\frac{2\pi}{\lambda_0}\Delta n(z)  + \beta_N(z,N,\Vert\Psi\Vert^2),\\
\beta_N(z,N,\Vert\Psi\Vert^2) &= \frac{2\pi}{\lambda_0}\Delta n_N(z,N,\Vert\Psi\Vert^2) + 
\frac{i}{2}\left[g(z,N,\Vert\Psi\Vert^2) - \alpha(z,N,\Vert\Psi\Vert^2)\right]. \label{eq:betaN}
\end{align}
\end{subequations}
The Langevin forces
\begin{equation*}
F_\mathrm{sp}(z,t,N)=
\begin{bmatrix}
F_\mathrm{sp}^+(z,t,N) \\
F_\mathrm{sp}^-(z,t,N)
\end{bmatrix},
\end{equation*}
model the spontaneous emission into the forward and backward traveling waves
and are assumed to have zero mean
\begin{equation*}
\langle F_\mathrm{sp}\rangle=0.
\end{equation*}
Here, $n_\mathrm{g}$ is the real--valued group index, $\beta$ is the propagation factor and $\Delta n$ is the  index detuning with respect to the reference modal index. $\beta_N$ describes the dependence of the propagation factor on the excess carrier density $N$, consisting of a real part proportional to $\Delta n_N$ and an imaginary part with $g$ being the modal gain and $\alpha$ the optical losses. $\kappa^\pm$
are complex coupling coefficients describing a Bragg grating
(coupling forward and backward traveling waves) and $\widehat{\cal D}(z,t)$ is the dispersion operator.
The ensemble or temporal average is denoted by $\langle \cdot \rangle$.
The dependence of $\Delta n_N$, $g$ and $\alpha$ on the optical power $\Vert\Psi\Vert^2$ is due to nonlinear effects like the Kerr effect, gain compression, and two-photon absorption, which affect the stationary states only slightly but can have a significant impact on the dynamic properties.

All quantities entering \eqref{eq:cwe} are obtained by weighting the original quantities entering Maxwell's equations with the intensity distribution $e^2(x,y)$  of the transverse mode.
If we choose the origin of $z$ such that the grating given by a complex dielectric function $\varepsilon(x,y,z)$ has the property $\varepsilon(x,y,z)=\varepsilon(x,y,-z)$, then $\kappa^+=\kappa^-=\kappa$ holds.
Eq. \eqref{eq:cwe} has to be solved subject to the usual boundary conditions
\begin{equation}
\label{eq:bc}
\Psi^+(0,t)=r_0\Psi^-(0,t)\quad\mathrm{and}\quad\Psi^-(L,t)=r_L\mathrm{e}^{-2i\beta_0L}\Psi^+(L,t),
\end{equation}
with $L$ being the total cavity length and $r_0$ and $r_L$ the complex--valued reflection coefficients at the facets at $z=0$ and $z=L$, respectively.
Scattering matrices establish transition conditions on $\Psi$ at the interfaces between different sections \cite{tromborg1994traveling, radziunas2017traveling}.

The coupled--wave equations \eqref{eq:cwe} must be supplemented with an equation for the (excess) carrier density $N$.
Although the theoretical treatment is quite general, we will specifically treat diode lasers in this paper, where the carrier dynamics is governed actually by both electrons and holes. 
We assume charge neutrality and consider only the excess carriers in the active region with density $N=n-n_0=p-p_0$ (where $n$ and $p$ are the electron and hole densities and $n_0$, $p_0$ are the corresponding equilibrium densities), neglecting any transport and capture effects.
Thus, the excess carrier density is assumed to obey the rate equation
\begin{equation}
\label{eq:carrier}
\frac{\partial N}{\partial t} - \frac{j(z,N)}{q d} + R(z,N) + R_\mathrm{st}(z,N,\Vert\Psi\Vert^2) = F_N(z,t,N)
\end{equation}
with the rate of spontaneous and non-radiative recombination $R$ and the
rate of stimulated recombination \cite{radziunas2017traveling}%
\footnote{It can be derived by multiplying \eqref{eq:cwe} from the left with $[\Psi^{+\ast},\Psi^{-\ast}]$, the complex conjugate of \eqref{eq:cwe} with $[\Psi^{+},\Psi^{-}]$ and adding both equations to obtain a balance equation for the electromagnetic energy in differential form (Poynting's theorem).}
\begin{equation}
\label{eq:rstim_orig}
R_\mathrm{st} = 
 \frac{\Re{ \left(\Psi^\ast \cdot [g -2i\widehat{\cal D}]\Psi\right) }}{dW\hbar\omega_0},
\end{equation}
where the dot product means $\Psi_1\cdot\Psi_2 = \Psi_1^+\Psi_2^++\Psi_1^-\Psi_2^-$.
Here, $q$ is the elementary charge, $\hbar$ the reduced Planck's constant and $d$ and $W$ are the thickness and width, respectively, of the active region.
The current density can be related to the (sectional) applied voltage by means of the Joyce model \cite{joyce1980current, zeghuzi2018modeling}
\begin{equation}
\label{eq:currentdensity}
j(z,N)=\frac{U_s-U_\mathrm{F}(N)}{WL_sR_s}
\end{equation}
with sectional applied voltage $U_s$,  resistance $R_s$,  injection current $I_s$,  length $L_s$, and the Fermi voltage $U_\mathrm{F}$. The carrier--density dependent injection current density given by \eqref{eq:currentdensity} counteracts longitudinal spatial hole burning, which substantially lowers the degradation of the side mode suppression in strongly coupled DFB lasers compared to a model assuming a constant current density \cite{bandelow1992influence}. The frequency modulation response, albeit not topic of the present paper, is also affected as shown in \cite{lassen1993influence}.

Dispersion, \textit{i.e.}, the frequency dependence of the dielectric function, enters \eqref{eq:cwe} first via the group index $n_\mathrm{g}$ in front of the time derivative resulting from a linearization of the real part of the dielectric function with respect to the frequency $\omega$. Second, the dispersion of the optical gain (and associated refractive index) is described in the time domain by the operator $\widehat{\cal D}(z,t)$, which can be obtained, \textit{e.g.}, by approximating the gain spectrum by a Lorentzian (or a series of Lorentzians) and a back-transformation into the time domain. This results in auxiliary differential equations for polarization functions \cite{ning1997effective, bandelow2001impact}, see Appendix~\ref{appendix:dispersion}. Due to the fact, that the spectral width of the gain in semiconductor lasers is much larger than the width of the cavity resonances,  $\widehat{\cal D}(z,t)$ can be expanded again linearly in the frequency domain. In contrast to \cite{ning1997effective}, we neglect here the dependence of the dispersion operator on the carrier density $N$.

Besides $\Psi$, we consider $N$ as the only further stochastic variable 
and spontaneous emission as the only source of noise, neglecting all other noise sources.
The theory can be easily extended to include carrier noise as sketched in the Outlook in Sec.~\ref{sec:outlook}.
We assume that $\kappa$ does not depend on $N$, \textit{i.e.}, we exclude gain--coupled lasers, and
neglect fluctuations of the shape of the power profile.
We neglect the impact of non-lasing side modes on the linewidth of the lasing mode and employ a single--mode approximation. 
Thus the linewidth rebroadening caused by mode partitioning and poor side--mode suppression, \textit{cf.} \cite{krueger1988semiconductor,pan1991linewidth}, can not be not accounted for.

Regarding the properties of the noise \cite{gardiner2009stochastic}, we employ the usual assumptions of ergodicity (ensemble or statistical average equals temporal average) and stationarity (\textit{i.e.}, $\langle \xi(t)\xi(t')\rangle$ depends only on the time difference $t-t'$) of any stochastic variable $\xi$. 
The fluctuations are assumed to have Gaussian probability distributions and delta--correlated covariance functions in time (Markov approximation) and space, \textit{i.e.}, $\langle \xi(z,t)\zeta(z',t')\rangle=2D_{\xi\zeta}(z)\delta(t-t')\delta(z-z')$ with $D_{\xi\zeta}$ being called diffusion coefficient.
In order to obtain a Lorentzian shape of the optical spectrum around the lasing frequency, some approximations have to be employed which are described in Sec.~\ref{sec:Lorentzian}. 
The full width at half maximum (FWHM) of the Lorentzian is commonly called \emph{intrinsic} linewidth. As we consider only spontaneous emission noise, we calculate the quantum limit of the intrinsic linewidth. 
If non-Markovinan noise is considered, which is beyond of the scope of the paper, the lineshape tends towards a Gaussian in the vicinity of the lasing frequency with a correspondingly larger FWHM \cite{agrawal1988effect, didomenico2010}.

\section{Equation of the field amplitude}
\label{sec:field}

Throughout this work, we restrict ourselves to the analysis of the noise at continuous wave (CW) emission, \textit{i.e.}, at steady-state lasing.
Therefore, we expand $\Psi(z,t)$ in terms of the stationary eigenmodes $\Phi_m(z)$ (longitudinal modes),
\begin{equation}
\label{eq:Ansatz}
\Psi(z,t)=\sum_mf_m(t)\Phi_m(z),
\end{equation}
satisfying the equation
\begin{equation}
\label{eq:Phi}
\left[\sigma_z \frac{\partial}{\partial z} + iM\left(z,\overline{N},\overline{\Vert\Psi\Vert^2}\right) + i\beta_{\cal D}(z,\Omega_m)
+i\frac{n_{\mathrm{g}}(z)}{c}\Omega_m\right]\Phi_m(z)
=0
\end{equation}
with
\begin{align}
M\left(z,\overline{N},\overline{\Vert\Psi\Vert^2}\right) &= \begin{bmatrix}
\Delta\beta\left(z,\overline{N},\overline{\Vert\Psi\Vert^2}\right)  &
\kappa(z) \\
\kappa(z) &
\Delta\beta\left(z,\overline{N},\overline{\Vert\Psi\Vert^2}\right)
\end{bmatrix}\equiv\overline{M}, \label{eq:matrixMbar}\\
\Delta\beta\left(z,\overline{N},\overline{\Vert\Psi\Vert^2}\right) &= \beta\left(z,\overline{N},\overline{\Vert\Psi\Vert^2}\right)-\beta_0\equiv\overline{\beta}-\beta_0=\Delta\overline{\beta}.
\label{eq:betaoverline}
\end{align}
Here, $\overline{N}$ and $\overline{\Vert\Psi\Vert^2}$ are the steady-state carrier density and optical power, respectively, at which the operator of the eigenvalue problem \eqref{eq:Phi} is expanded ($\overline{\beta}$, $\overline{M}$ are the corresponding propagation factor and matrix $M$).
The complex eigenvalue $\Omega_m$ describes both, the frequency deviation 
$\Re{(\Omega_m)}=\omega_m-\omega_0$ (or the wavelength
deviation $\Delta\lambda_m=\Re(\Omega_m)\, d\lambda/d\omega\vert_{\lambda_0}$), and the 
mode damping rate $\Im{(\Omega_m)}$.
The dispersion $\beta_{\cal D}(z,\Omega_m)$ is given by the operator
$\mathcal{D}(z,t)$ in the frequency domain.
Eq.~\eqref{eq:Phi} has to be solved subject to the boundary conditions
\begin{equation}
\label{eq:bc2}
\begin{aligned}
\Phi_m^+(0)&=r_0\Phi_m^-(0)\\
\Phi_m^-(L)&=r_L\mathrm{e}^{-2i\beta_0L}\Phi_m^+(L)
\end{aligned}
\end{equation}
following from \eqref{eq:bc}.

In the case of stationary eigenmodes considered here, the two-point boundary value problem \eqref{eq:Phi} together with the boundary conditions \eqref{eq:bc2} is in fact a quasi-linear eigenvalue problem. The  point of expansion $\big(\overline{\Vert\Psi\Vert^2},\overline{N}\big)$ of the nonlinear operator must be chosen self-consistently, \textit{i.e.}, it is required to satisfy the conditions \eqref{eq:threshold} and \eqref{eq:carrier_steady} for the (mean) stationary state given below.
Furthermore, in the presence of dispersion $\beta_{\mathcal{D}}(\Omega_m)$, the eigenvalue problem is also nonlinear in the eigenvalue.
The problem is similar to the \emph{self-consistent field method} in electronic structure theory \cite{woods2019} and can be treated in a similar way (e.g., linearization of the optical-power and the frequency dependency of the gain dispersion and fixed-point iteration, where the point of expansion is updated to the most recent stationary state in each step).
Alternatively, the system can be directly propagated in the time-domain (as it has been done in Sec.~\ref{sec:example}), until convergence to a suitable CW state has been achieved.

Note that different modes (eigensolutions of \eqref{eq:Phi}) fulfill the orthogonality relation%
\footnote{It can be derived by multiplying \eqref{eq:Phi} from the left by $[\Phi^-_n,\Phi^+_n]$, the corresponding equation for $\Phi_n$ by $[\Phi^-_m,\Phi^+_m]$, integrating along $z$, using \eqref{eq:bc2} and subtracting both equations.}
\begin{equation}
\label{eq:orthogon}
(\Phi_m,\frac{\partial \beta}{\partial \omega}\Phi_n)=0\quad\mathrm{for}\quad m\neq n
\end{equation}
with the inner product (distinguished from a standard Hilbert space scalar product)
\begin{equation}
\label{eq:inner_product}
(\Phi,\Psi)=\int_0^L(\Phi^+\Psi^- + \Phi^-\Psi^+)\,dz.
\end{equation}
At an exceptional point the expansion \eqref{eq:Ansatz} includes also a generalized eigenmode \cite{wenzel1996mechanisms}.
For $m=n$, the factor $\partial \beta/\partial \omega$, which can be interpreted as an inverse complex--valued group velocity, is the exact derivative 
\begin{equation}
\label{eq:dbetadeomega}
\frac{\partial\beta}{\partial \omega} = 
\frac{n_\mathrm{g}}{c}+
\frac{\partial\beta_{\cal D}}{\partial \omega}\Big|_{\Omega_m}.
\end{equation}
In the general case of $m\neq n$ this holds only approximately:
\begin{equation}
\label{eq:dbetadomega}
\frac{\partial \beta}{\partial \omega}
=\frac{n_\mathrm{g}}{c}+
\frac{\beta_{\cal D}(\Omega_m)-\beta_{\cal D}(\Omega_n)}{\Omega_m-\Omega_n}
\approx
\frac{n_\mathrm{g}}{c}+
\frac{\partial\beta_{\cal D}}{\partial\omega}\Big|_{\Omega_m}.
\end{equation}

A system of ordinary differential equations for the amplitudes $f_m(t)$ can be obtained by inserting \eqref{eq:Ansatz} into \eqref{eq:cwe}, using \eqref{eq:Phi}, multiplying from left with $[\Phi^-_n,\Phi^+_n]$, and integrating along $z$.
In order to exploit the orthogonality relation \eqref{eq:orthogon} with the approximation \eqref{eq:dbetadomega}, we have again to expand
\begin{equation}
\beta_{\cal D}(\tilde{\omega})-\beta_{\cal D}(\Omega_m)=\frac{\partial\beta_{\cal D}}{\partial\omega}\Big|_{\Omega_m}(\tilde{\omega}-\Omega_m),
\end{equation}
where $\tilde{\omega}=\omega-\omega_0$.
Note that this is consistent with the slowly varying amplitude approximation.
Transforming back into the time domain yields
\begin{equation}
\int\left[\beta_{\cal D}(\tilde{\omega})-\beta_{\cal D}(\Omega_m)\right]f_m(\tilde{\omega})\mathrm{e}^{i\tilde{\omega} t}\,d\tilde{\omega}
=\left[\widehat{\cal D}-\beta_{\cal D}(\Omega_m)\right]f_m(t)
\end{equation}
and
\begin{equation}
\begin{aligned}
\frac{\partial\beta_{\cal D}}{\partial\omega}\Big|_{\Omega_m}
\int (\tilde{\omega}-\Omega_m) f_m(\tilde{\omega}) \mathrm{e}^{i\tilde{\omega} t}\,d\tilde{\omega}
&=-\frac{\partial\beta_{\cal D}}{\partial\omega}\Big|_{\Omega_m}
\left[\Omega_m f_m(t)
+i\frac{\partial}{\partial t}\int f_m(\tilde{\omega}) \mathrm{e}^{i\tilde{\omega} t}\,d\tilde{\omega}\right]\\
&=-\frac{\partial\beta_{\cal D}}{\partial\omega}\Big|_{\Omega_m}
\left[\Omega_m f_m(t)
+i\frac{\partial f_m}{\partial t}\right].
\end{aligned}
\end{equation}
Therefore the amplitudes $f_m$ fulfill
\begin{equation}
\label{eq:f}
(\Phi_m,\frac{\partial \beta}{\partial \omega}\Phi_m)\frac{\partial f_m}{\partial t} - i\Omega_m(\Phi_m, \frac{\partial \beta}{\partial \omega}\Phi_m)f_m
+ i\sum_n(\Phi_m,\Delta M\Phi_n)f_n
= (\Phi_m,F_\mathrm{sp})
\end{equation}
with 
\begin{equation}
\label{eq:DeltaM}
\Delta M = M-\overline{M}=(\beta-\overline{\beta})
\begin{bmatrix}
1  &
0 \\
0 &
1
\end{bmatrix}.
\end{equation}

In what follows, we employ the single mode approximation and drop the subscript $m$.
Then Eq. \eqref{eq:f} becomes a single differential equation for the complex--valued amplitude $f$ of the lasing mode
\begin{equation}
\label{eq:fnew}
\frac{\partial f}{\partial t} - i\Omega f + i\frac{(\Phi,\Delta M\Phi)}{(\Phi,\frac{\partial \beta}{\partial \omega}\Phi)}f 
 = F_f.
\end{equation}
The new Langevin noise source entering \eqref{eq:fnew} is given by
\begin{equation}
\label{eq:Ff}
F_f=\frac{(\Phi,F_\mathrm{sp})}{(\Phi,\frac{\partial \beta}{\partial \omega}\Phi)}.
\end{equation}
For an analysis above threshold, it is beneficial to derive equations for the modulus $|f|$ and the phase $\varphi$ of $f$, because the fluctuations of $|f|$ are damped but those of $\varphi$ are not.
For the modulus squared $|f|^2$ (called intensity in the rest of the paper) it follows
\begin{equation}
\label{eq:modulus}
\frac{\partial |f|^2}{\partial t}
+ 2\Im{(\Omega)}|f|^2
 -2\Im{\frac{(\Phi,\Delta M\Phi)}{(\Phi,\frac{\partial \beta}{\partial \omega}\Phi)} }|f|^2
= 2\Re{(F_ff^\ast)}.
\end{equation}
An equation for the phase can be determined from 
\begin{equation}
\label{eq:Im}
\Im{\left(\frac{\partial f}{\partial t}f^\ast\right)}
-\Re{(\Omega)}|f|^2
 + \Re{ \frac{(\Phi,\Delta M\Phi)}{(\Phi,\frac{\partial \beta}{\partial \omega}\Phi)}} |f|^2
= \Im{(F_ff^\ast)}
\end{equation}
and
\begin{equation}
\frac{\partial f}{\partial t}
=\mathrm{e}^{i\varphi}\frac{\partial |f|}{\partial t}+ if \frac{\partial \varphi}{\partial t}\quad\Longrightarrow\quad 
\Im{\left(\frac{\partial f}{\partial t}f^\ast\right)} = |f|^2 \frac{\partial \varphi}{\partial t}.
\end{equation}
The stochastic term on the right--hand side of \eqref{eq:modulus} describes intensity noise, which does not have a vanishing expectation value anymore: $\langle F_f f^\ast \rangle \neq 0$. It is therefore convenient to rewrite the term by distilling out the expectation value and introducing two new real-valued Langevin forces $F_{|f|^2}$ and $F_\varphi$ with zero mean as
\begin{equation}
\label{eq:Langevinnew}
2F_ff^\ast=2\langle F_ff^\ast\rangle+F_{|f|^2}+2i|f|^2F_\varphi.
\end{equation}
It will be shown below that $\langle F_ff^\ast\rangle$ is real--valued, see Eq.~\eqref{eq:Fff}.
Then, the final equations for the modulus squared and phase of $f$ read
\begin{equation}
\label{eq:amplitude}
\frac{\partial |f|^2}{\partial t}
=- 2\Im{(\Omega)}|f|^2 
+2\Im{\frac{(\Phi,\Delta M\Phi)}{(\Phi,\frac{\partial \beta}{\partial \omega}\Phi)} }|f|^2 
+ 2\langle F_ff^\ast\rangle+F_{|f|^2}
\end{equation}
and
\begin{equation}
\frac{\partial \varphi}{\partial t}
= \Re{(\Omega)}
- \Re{\frac{(\Phi,\Delta M\Phi)}{(\Phi,\frac{\partial \beta}{\partial \omega}\Phi)}} 
+ F_\varphi.
\end{equation}
Now we consider small fluctuations
\begin{align}
\label{eq:variationN}
N(z,t) &=  \langle N(z)\rangle +\delta N(z,t), \\
\label{eq:variationphi}
\varphi(t)  & =\langle\varphi(t)\rangle+\delta\varphi(t),\\
\label{eq:variationfb2}
|f(t)|^2 & =\langle|f|^2\rangle+\delta|f(t)|^2 
\end{align}
and expand
\begin{equation}
\label{eq:variationbeta}
\beta = \beta(\langle N\rangle, \langle |f|^2\rangle) + \frac{\partial \beta}{\partial N}\bigg|_{\langle N\rangle,\langle |f|^2\rangle}\delta N 
 +\frac{\partial \beta}{\partial |f|^2}\bigg|_{\langle N\rangle,\langle|f|^2\rangle}\delta |f|^2 
\end{equation}
around the mean values
\begin{equation}
\label{eq:meanvalues}
\begin{aligned}
\langle N(z)\rangle &= \overline{N}(z), \\
\langle\varphi(t)\rangle & =\Re{(\Omega)} t.
\end{aligned}
\end{equation}
The mean intensity $\langle|f|^2\rangle$ is required to satisfy
\begin{equation}
\label{eq:mean}
\Im{(\Omega)}\langle|f|^2\rangle
= \langle F_ff^\ast\rangle.
\end{equation}
Due to the extreme smallness of the spontaneous emission $\langle F_ff^\ast\rangle$ going into the lasing mode, 
\eqref{eq:mean} can be replaced by
\begin{equation}
\label{eq:threshold}
\Im{(\Omega)}=0
\end{equation}
above threshold $(\langle|f|^2\rangle>0)$, which will be exploited in what follows (\textit{i.e.}, $\Omega=\Re{(\Omega)}$).
The intensity and phase fluctuations fulfill the equations
\begin{equation}
\label{eq:fbetrag}
\frac{\partial \delta |f|^2}{\partial t}
= 2\Im{ \frac{(\Phi,\frac{\partial \beta}{\partial N}\delta N\Phi)+(\Phi,\frac{\partial \beta}{\partial |f|^2}\Phi)\delta |f|^2}
{(\Phi,\frac{\partial \beta}{\partial \omega}\Phi)} } \langle|f|^2\rangle + F_{|f|^2}
\end{equation}
and
\begin{equation}
\label{eq:phase}
\frac{\partial \delta\varphi}{\partial t}
= -\Re{ \frac{(\Phi,\frac{\partial \beta}{\partial N}\delta N\Phi)+(\Phi,\frac{\partial \beta}{\partial |f|^2}\Phi)\delta |f|^2}{(\Phi,\frac{\partial \beta}{\partial \omega}\Phi)} } + F_\varphi,
\end{equation}
which are the basis for the derivation of the linewidth formula in the following sections.

\section{Lorentzian line shape}
\label{sec:Lorentzian}

The power spectral density (PSD) of the optical field at $z=0$ can be calculated utilizing the Wiener--Khinchin theorem as\footnote{We do not address technical issues regarding the non-existence of Fourier integrals of fluctuating quantities here and refer to Ref.~\cite{landau1980statistical}.}
\begin{equation}
S_{E}(\omega)=\iint\int_{-\infty}^{+\infty}\left\langle E^\ast(x,y,0,t) E(x,y,0,t+\tau)\right\rangle\mathrm{e}^{-i\omega \tau}\,d\tau\, dx dy
\end{equation}
with
\begin{equation}
\langle E^\ast(x,y,0,t) E(x,y,0,t+\tau)\rangle=\frac{1-|r_0|^2}{2\varepsilon_0cn}e^2(x,y)\langle f^\ast(t)f(t+\tau)\rangle|\Phi^-(0)|^2\mathrm{e}^{i\omega_0\tau} + \text{c.c.}
\end{equation}
Hence we have to calculate the spectral density 
\begin{equation}
S_{f}(\omega)=\int_{-\infty}^{+\infty}\left\langle f^\ast(t) f(t+\tau)\right\rangle \mathrm{e}^{-i(\omega-\omega_0)\tau}\,d\tau
\end{equation}
with
\begin{equation}
\label{eq:ffcorrel}
\langle f^\ast(t)f(t+\tau)\rangle 
= \langle |f(t)||f(t+\tau)|\mathrm{e}^{i[\varphi(t+\tau)-\varphi(t)]}\rangle.
\end{equation}
Now we employ here a first approximation: We neglect in \eqref{eq:ffcorrel} the intensity fluctuations because above threshold they are damped, in contrast to the phase fluctuations, which can been seen by comparing \eqref{eq:fbetrag} and \eqref{eq:phase}, see the discussion in \cite{vahala1983semiclassical}.
Therefore, we obtain
\begin{equation}
\label{eq:ff}
\langle f^\ast(t)f(t+\tau)\rangle 
\approx \langle|f|^2\rangle\langle\mathrm{e}^{i\Delta_\tau\varphi}\rangle\mathrm{e}^{i\Omega\tau}
\end{equation}
with
\begin{equation}
\Delta_\tau\varphi=\delta\varphi(t+\tau)-\delta\varphi(t)
\end{equation}
and $\Omega=\Re{(\Omega)}$.
The random variable $\Delta_\tau\varphi$ is assumed to be Gaussian distributed \cite{henry1986phase}, such that it holds
\begin{equation}
\label{eq:Gaussian}
\langle\mathrm{e}^{i\Delta_\tau\varphi}\rangle
=  \frac{1}{\sqrt{2\pi\mathop{\mathrm{Var}}(\Delta_\tau\varphi)}}\int_{-\infty}^{\infty}\mathrm{e}^{-\frac{(\Delta_\tau\varphi)^2}
{2\mathop{\mathrm{Var}}(\Delta_\tau\varphi)}}
\mathrm{e}^{i\Delta_\tau\varphi}d\Delta_\tau\varphi
= \mathrm{e}^{-\frac{1}{2}\mathop{\mathrm{Var}}(\Delta_\tau\varphi)}.
\end{equation}
The variance of the phase fluctuations is
\begin{equation}
\begin{aligned}
\mathop{\mathrm{Var}}(\Delta_\tau\varphi)
&=  \langle\left(\Delta_\tau\varphi-\langle\Delta_\tau\varphi\rangle\right)^2\rangle\\
&=  \langle(\Delta_\tau\varphi)^2\rangle\\
&=  \langle \left(\delta\varphi(t+\tau)\right)^2 \rangle - 2\langle\delta\varphi(t)\delta\varphi(t+\tau)\rangle + \langle \left( \delta\varphi(t) \right)^2\rangle\\
&=  2\left[\langle\left(\delta\varphi(0)\right)^2\rangle-\langle\delta\varphi(0)\delta\varphi(\tau)\rangle\right],
\end{aligned}
\end{equation}
where we exploited $\langle\Delta_\tau\varphi\rangle=0$ and the stationarity property. 
We introduce the PSD of phase fluctuations\footnote{Note that the spectral density of phase fluctuations is even in the frequency $S_{\delta\varphi}(\tilde{\omega})=S_{\delta\varphi}(-\tilde{\omega})$.}
\begin{equation}
\label{eq:Sphi}
S_{\delta\varphi}(\tilde{\omega})=\int_{-\infty}^{+\infty}\left\langle\delta\varphi(t)\delta\varphi(t+\tau)\right\rangle\mathrm{e}^{-i\tilde{\omega}\tau}\,d\tau
\end{equation}
and the PSD of optical frequency fluctuations 
$\delta\omega(\tilde{\omega})=i\tilde{\omega}\delta\varphi(\tilde{\omega})$
\begin{equation}
\label{eq:Somega}
S_{\delta\omega}(\tilde{\omega})=\int_{-\infty}^{+\infty}\left\langle\delta\omega(t)\delta\omega(t+\tau)\right\rangle\mathrm{e}^{-i\tilde{\omega}\tau}\,d\tau=\tilde{\omega}^2S_{\delta\varphi}(\tilde{\omega}).
\end{equation}
With these spectral densities, the variance of the phase fluctuations can be written as
\begin{equation}
\begin{aligned}
\mathop{\mathrm{Var}}(\Delta_\tau\varphi)
&= \frac{1}{\pi}\int_{-\infty}^{\infty}S_{\delta\varphi}(\tilde{\omega})\left(1-\mathrm{e}^{i\tilde{\omega} \tau}\right)\,d\tilde{\omega}\\
&= \frac{1}{\pi}\int_{-\infty}^{\infty}S_{\delta\varphi}(\tilde{\omega})\left(2\sin^2\left(\frac{\tilde{\omega}\tau}{2}\right)-i\sin\left(\tilde{\omega}\tau\right)\right)\,d\tilde{\omega}\\
&=  \frac{2}{\pi}\int_{-\infty}^{\infty}S_{\delta\varphi}(\tilde{\omega})\sin^2\left(\frac{\tilde{\omega}\tau}{2} \right) \,d\tilde{\omega}\\
&= \frac{\tau^2}{2\pi}\int_{-\infty}^{\infty}S_{\delta\omega}(\tilde{\omega}) \left( \frac{\sin\left(\frac{\tilde{\omega}\tau}{2}\right) }{\frac{\tilde{\omega}\tau}{2}} \right)^2 \,d\tilde{\omega}.
\end{aligned}
\label{eq:Var_deltaphi_1}
\end{equation}
We carry out a second approximation by noting that the function $\sin^2\left(\frac{\tilde{\omega}\tau}{2}\right)/(\frac{\tilde{\omega}\tau}{2})^2$ peaks strongly at $\tilde{\omega}=0$ for large $|\tau|$, and obtain
\begin{equation}
\label{eq:Varphi}
\mathop{\mathrm{Var}}(\Delta_\tau\varphi)
\approx \frac{|\tau|}{\pi} S_{\delta\omega}(0)  \int_{-\infty}^{\infty}\frac{\sin^2x}{x^2}\,dx
= |\tau| \, S_{\delta\omega}(0).
\end{equation}
In this step, we have recovered a central property of Brownian motion, as we can observe that the root mean square displacement of the phase fluctuation grows as $\propto\sqrt{|\tau|}$.
Inserting \eqref{eq:Varphi} into \eqref{eq:Gaussian} and, finally, the result into \eqref{eq:ff} yields
\begin{equation}
\langle f^\ast(t)f(t+\tau)\rangle 
\approx  \langle|f|^2\rangle\mathrm{e}^{-\frac{1}{2} S_{\delta\omega}(0)|\tau| }\mathrm{e}^{i\Omega\tau}
\end{equation}
and
\begin{equation}
\begin{aligned}
S_{f}(\omega)
&\approx\langle|f|^2\rangle\int_{-\infty}^{+\infty} \mathrm{e}^{-\frac{1}{2}S_{\delta\omega}(0)|\tau|+i(\omega_0+\Omega-\omega)\tau}\,d\tau\\
&=  \langle|f|^2\rangle  \frac{  S_{\delta\omega}(0)}{\left(\omega - (\omega_0+\Omega)\right)^2 + \left( \frac{S_{\delta\omega}(0)}{2}  \right)^2}.
\end{aligned}
\end{equation}
This is a Lorentzian centered at $\omega=\omega_0+\Omega$ with the FWHM
\begin{equation}
\label{eq:FWHM}
\Delta\omega=S_{\delta\omega}(0)
\end{equation}
given by the PSD of the optical frequency fluctuations taken at zero Fourier frequency.

In the following sections of this paper, we will focus on the static white noise limit, where the frequency noise PSD $S_{\delta \omega}(\omega)$ is effectively approximated by a constant, such that the second approximation carried out on the step from Eq.~\eqref{eq:Var_deltaphi_1} to \eqref{eq:Varphi} becomes in fact redundant.

\section{Correlation functions}
\label{sec:correl}

The correlation function of the spontaneous emission noise $F_f$ defined in \eqref{eq:Ff} is
\begin{equation}
\langle F_{f}^\ast(t)F_{f}(t')\rangle
=\frac{\langle(\Phi^\ast,F_\mathrm{sp}^\ast(z,t))(\Phi,F_\mathrm{sp}(z',t'))\rangle}{\left|(\Phi,\frac{\partial \beta}{\partial \omega}\Phi)\right|^2}.
\end{equation}
The numerator reads
\begin{equation}
\label{eq:correlation_function}
\begin{aligned}
&\langle(\Phi^\ast,F_\mathrm{sp}^\ast(z,t))(\Phi,F_\mathrm{sp}(z',t'))\rangle\\
&=\int_0^L\int_0^L
\left\langle\left[
\Phi^{+\ast}F_\mathrm{sp}^{-\ast}(z,t)+\Phi^{-\ast}F_\mathrm{sp}^{+\ast}(z,t)
\right]
\left[
\Phi^{+}F_\mathrm{sp}^{-}(z',t')+\Phi^{-}F_\mathrm{sp}^{+}(z',t')
\right]\right\rangle
\,dzdz'\\
&=\int_0^L\int_0^L\left[
|\Phi^+|^2 \langle F_\mathrm{sp}^{-\ast}(z,t)F_\mathrm{sp}^{-}(z',t')\rangle
+|\Phi^-|^2 \langle F_\mathrm{sp}^{+\ast}(z,t)F_\mathrm{sp}^{+}(z',t')\rangle
\right]\,dzdz'\\
&=2\int_0^L \Vert\Phi(z)\Vert^2D_{\mathrm{sp}}(z)\,dz \, \delta(t-t^\prime)
\end{aligned}
\end{equation}
with 
\begin{equation}
\Vert\Phi\Vert^2=|\Phi^+|^2+|\Phi^-|^2
\end{equation}
taking into account 
the noise covariance functions \cite{marani1995spontaneous,henry1996quantum,henry1986theory} for index-- and absorption coupling\footnote{For gain--coupling see \cite{baets1993distinctive}.}
\begin{equation}
\begin{aligned}
\langle F_\mathrm{sp}^{\pm\ast}(z,t) F_\mathrm{sp}^{\pm}(z',t') \rangle 
&=2D_{\mathrm{sp}}(z)\delta(t-t^\prime)\delta(z-z^\prime),\\
\langle F_\mathrm{sp}^{\mp\ast}(z,t) F_\mathrm{sp}^{\pm}(z',t') \rangle 
&=0,\\
\langle F_\mathrm{sp}^{\pm}(z,t) F_\mathrm{sp}^{\pm}(z',t') \rangle 
&=0,\\
\langle F_\mathrm{sp}^{\mp}(z,t) F_\mathrm{sp}^{\pm}(z',t') \rangle 
&=0
\end{aligned}
\end{equation}
with
\begin{equation}
\label{eq:Dsp}
2D_\mathrm{sp}(z,N,\Vert \Psi \Vert^2 )=\hbar\omega_0 n_\mathrm{sp}(N) g(z,N, \Vert \Psi \Vert^2 ),
\end{equation}
where $n_\mathrm{sp}$ is the population inversion factor (a Bose--Einstein distribution function multiplied by $-1$).
Since $\Psi$ has the unit $\sqrt{\text{W}}$, $D_\mathrm{sp}$ must have the unit Ws/m as it is.
The diffusion coefficient $D_{f^\ast f}$ defined by
\begin{equation}
\label{eq:FFcorrelationfunction}
\langle F_{f}^\ast(t)F_{f}(t')\rangle
=2D_{f^\ast f}\delta(t-t^\prime).
\end{equation}
is therefore
\begin{equation}
\label{eq:Dff}
D_{f^\ast f} =D_{f^\ast f}(N,\Vert \Psi \Vert^2)
=\frac{\int_0^L \Vert\Phi(z)\Vert^2 D_{\mathrm{sp}}(z,N,\Vert \Psi \Vert^2 )\,dz}{\left|(\Phi,\frac{\partial \beta}{\partial \omega}\Phi)\right|^2}.
\end{equation}
The correlation functions of the Langevin forces entering \eqref{eq:fbetrag} and \eqref{eq:phase}  read
\begin{align}
\label{eq:f2f2}
\langle F_{|f|^2}(t)F_{|f|^2}(t^\prime)\rangle&=2D_{|f|^2|f|^2}\delta(t-t^\prime)=4D_{f^\ast f}\langle|f|^2\rangle\delta(t-t^\prime),\\
\label{eq:phiphi}
\langle F_\varphi(t) F_\varphi(t^\prime)\rangle&=2D_{\varphi\varphi}\delta(t-t^\prime)=\frac{D_{f^\ast f}}{\langle|f|^2\rangle}\delta(t-t^\prime),\\
\label{eq:f2phi}
\langle F_{|f|^2}(t)F_\varphi(t^\prime)\rangle&=2D_{|f|^2\varphi}\delta(t-t^\prime)=0,
\end{align}
which can be derived using the transformation rules for the diffusion coefficients \cite{lax1966classical}
\begin{equation}
\begin{aligned}
D_{|f|^2|f|^2}&=\left(\frac{\partial |f|^2}{\partial f}\frac{\partial |f|^2}{\partial f^\ast}+\frac{\partial |f|^2}{\partial f^\ast}\frac{\partial |f|^2}{\partial f}\right)D_{f^\ast f},\\
D_{\varphi\varphi}&=\left(\frac{\partial \varphi}{\partial f}\frac{\partial \varphi}{\partial f^\ast}+\frac{\partial \varphi}{\partial f^\ast}\frac{\partial \varphi}{\partial f}\right)D_{f^\ast f},\\
D_{|f|^2\varphi}&=\left(\frac{\partial |f|^2}{\partial f}\frac{\partial \varphi}{\partial f^\ast}+\frac{\partial |f|^2}{\partial f^\ast}\frac{\partial \varphi}{\partial f}\right)D_{f^\ast f},
\end{aligned}
\end{equation}
and the relations $|f|^2=f^\ast f$ and $\varphi=-\frac{i}{2}\mathrm{ln}(f/f^\ast)$.
Following \cite{chow1994semiconductor}, the ensemble average defined in \eqref{eq:Langevinnew} can be obtained using the formal solution of \eqref{eq:fnew} in the form of
\begin{equation} 
\label{eq:formalsolution_f}
f\left(t\right)=f\left(t-\Delta t\right)+\int_{t-\Delta t}^{t}\partial f^{*}(t')/\partial t' \,dt',
\end{equation}
where the time step $0<\Delta t\ll\gamma_{R}^{-1}$ is assumed to be  short in comparison to the inverse relaxation rate $\gamma_{R}=\Im{\left(\Omega-\left(\Phi,\Delta M\Phi\right)/\left(\Phi,\partial_{\omega}\beta\Phi\right)\right)}$.
Substituting \eqref{eq:formalsolution_f} into $\langle F_{f}\left(t\right)f^{*}\left(t\right)\rangle$ yields
\begin{align*}
\langle F_{f}\left(t\right)f^{*}\left(t\right)\rangle	&= \langle F_{f}\left(t\right)f^{*}\left(t-\Delta t\right)\rangle + \int_{t-\Delta t}^{t}\langle F_{f}\left(t\right)\partial f^{*}(t')/\partial t'\rangle \,dt'\\
	&=\langle F_{f}\left(t\right)\rangle\langle f^{*}\left(t-\Delta t\right)\rangle-i\int_{t-\Delta t}^{t}\bigg(\Omega-\frac{\left(\Phi,\Delta M\Phi\right)}{\left(\Phi,\partial\beta/\partial\omega\,\Phi\right)}\bigg)^{*}\langle F_{f}\left(t\right)f^{*}\left(t'\right)\rangle \,dt'\\
	&\phantom{=}+\int_{t-\Delta t}^{t}\langle F_{f}\left(t\right)F_{f}^{*}\left(t'\right)\rangle\,dt',
\end{align*}
where we have used \eqref{eq:fnew} in the second step. The first term vanishes because of causality, since the fluctuations of $f^{*}\left(t-\Delta t\right)$ are not correlated with the future noise $F_{f}\left(t\right)$. Consequently, the correlation function factorizes and results in zero due to the zero-mean property of the Langevin force. The second term can be dropped since it is non-zero only over a set of measure zero (\emph{i.e.}, at $t=t'$). Finally, we obtain with \eqref{eq:FFcorrelationfunction}
\begin{equation}
\label{eq:Fff}
\langle F_{f}\left(t\right)f^{*}\left(t\right)\rangle	=\int_{t-\Delta t}^{t}\langle F_{f}\left(t\right)F_{f}^{*}\left(t'\right)\rangle\,dt'=D_{f^{*}f},
\end{equation}
where a factor 2 is canceled by 1/2 encountered in integrating only half of the $\delta$--function.

\section{Effective linewidth enhancement factor}

Using the ansatz \eqref{eq:Ansatz}, we can approximate the rate of stimulated emission \eqref{eq:rstim_orig} as
\begin{equation}
\label{eq:rstim}
R_\mathrm{st} = \frac{g+2\Im{\left(\beta_{\cal D}(\Omega)\right)}}{dW\hbar\omega_0} |f|^2 \Vert\Phi\Vert^2.
\end{equation}
The stationary mean value of the carrier density and its fluctuations satisfy
\begin{equation}
\label{eq:carrier_steady}
\frac{j(\langle N\rangle)}{ed} = R(\langle N\rangle) 
+ \frac{g(\langle N\rangle,\langle|f|^2\rangle
\Vert\Phi\Vert^2) + 2\Im{\left(\beta_{\cal D}(\Omega)\right)}}{dW\hbar\omega_0} \langle|f|^2\rangle
\Vert\Phi\Vert^2
\end{equation}
and
\begin{equation}
\label{eq:deltaRstim}
\frac{\partial \delta N}{\partial t}
+ \frac{\delta N}{\tau_\mathrm{d}} 
+ \frac{\partial R_\mathrm{st}}{\partial |f|^2} \delta|f|^2
= F_N,
\end{equation}
respectively, with the inverse differential carrier lifetime 
(taken at $\langle N \rangle$ and $\langle |f|^2 \rangle$) 
\begin{equation}
\frac{1}{\tau_\mathrm{d}}=
-\frac{1}{q d}\frac{\partial j}{\partial N}
+\frac{\partial R}{\partial N}
+\frac{\partial R_\mathrm{st}}{\partial N}.
\end{equation}
The FWHM of the Lorentzian line shape is given by the spectral density of the frequency fluctuations at zero Fourier frequency, \textit{cf.} \eqref{eq:FWHM}. This is equivalent to setting the time derivatives of the fluctuations equal to zero (static limit)
\begin{equation}
\frac{\partial \delta |f|^2}{\partial t}=0
\end{equation}
and
\begin{equation}
\frac{\partial \delta N}{\partial t} = 0.
\end{equation}
Furthermore, we omit carrier noise ($F_N=0$), considering only spontaneous emission noise as mentioned in Sec. \ref{sec:2}.
With these approximations, the carrier density fluctuations are directly related to the intensity fluctuations by
\begin{equation}
\label{eq:deltaN}
\delta N= - \tau_\mathrm{d}\frac{\partial R_\mathrm{st}}{\partial |f|^2}\delta|f|^2.
\end{equation}
We substitute \eqref{eq:deltaN} into \eqref{eq:fbetrag} and obtain
\begin{equation}
\label{eq:deltaf}
\delta|f|^2=\frac{F_{|f|^2}}{2\langle|f|^2\rangle\Im{(h_\alpha)}}
\end{equation}
with
\begin{equation}
\label{eq:halpha}
h_\alpha=\frac{(\Phi,\tau_\mathrm{d}\frac{\partial \beta}{\partial N}\frac{\partial R_\mathrm{st}}{\partial |f|^2} \Phi)
-(\Phi,\frac{\partial \beta}{\partial |f|^2}\Phi)}
{(\Phi,\frac{\partial \beta}{\partial \omega}\Phi)}.
\end{equation}
Inserting \eqref{eq:deltaf} back into \eqref{eq:deltaN} yields
\begin{equation}
\label{eq:deltaN2}
\delta N= - \tau_\mathrm{d}\frac{\partial R_\mathrm{st}}{\partial |f|^2}
\frac{F_{|f|^2}}{2\langle|f|^2\rangle\Im{(h_\alpha)}}.
\end{equation}
Finally, substituting \eqref{eq:deltaN2} into \eqref{eq:phase} results in
\begin{equation}
\label{eq:phase2}
\frac{\partial \delta\varphi}{\partial t}
= \frac{F_{|f|^2}}{2\langle|f|^2\rangle}\frac{\Re{(h_\alpha)}}{\Im{(h_\alpha)}}+ F_\varphi,
\end{equation}
which can be considered as a defining equation for the so--called effective Henry's $\alpha$-factor 
\begin{equation}
\label{eq:alphaHeff}
\alpha_\mathrm{H,eff}
= -\frac{\Re{(h_\alpha)}}{\Im{(h_\alpha)}}.
\end{equation}
If $\alpha_\mathrm{H,eff}\ne 0$, the Langevin force resulting in fluctuations of the intensity leads also to fluctuations of the phase.
By virtue of the different integrals in the numerator and denominator of \eqref{eq:halpha}, the carrier dependence of the propagation factor, nonlinear gain (gain compression), absorption (two--photon absorption) and index (Kerr effect) as well as gain dispersion are accounted for. The impact of longitudinal spatial hole burning and multi--section cavity structures (\textit{e.g.}, including Bragg gratings, passive sections or external cavities) is also correctly described.
If all of these additional effects are neglected, for a Fabry--P\'{e}rot (FP) laser the $\alpha$--factor is recovered
\begin{equation}
\label{eq:alfaH}
\alpha_\mathrm{H}=-\frac{\Re{\left(\frac{\partial \beta}{\partial N}\right)}}{\Im{\left(\frac{\partial \beta}{\partial N}\right)}}
=-\frac{4\pi}{\lambda_0}\frac{
\frac{\partial \Delta n_N}{\partial N}}
{\frac{\partial g}{\partial N}-\frac{\partial \alpha}{\partial N}}.
\end{equation}
The second equality is obtained using \eqref{eq:betaN}.
The minus sign in \eqref{eq:alphaHeff}, resulting in a corresponding minus sign in \eqref{eq:alfaH}, has been chosen in agreement with the usual convention  ensuring positive values of $\alpha_\mathrm{H}$ around the gain peak \cite{osinski1987linewidth}.
The derivative $\partial \alpha/\partial N$ of the modal absorption in \eqref{eq:alfaH} is often neglected, \textit{cf.} \cite{osinski1987linewidth}. 
One should keep in mind that $\alpha_\mathrm{H}$ is not a constant but depends on the carrier density (and the wavelength) because modal index and gain vary differently with carrier density.

\section{Spectral linewidth}
\label{sec:liniewidth}
Collecting the results of the previous sections, we can now calculate the spectral linewidth.
Rewriting \eqref{eq:FWHM} as
\begin{equation}
\Delta\omega=\int_{-\infty}^{+\infty}\left\langle\frac{\partial\delta\varphi(t)}{\partial t}
\frac{\partial\delta\varphi(t+\tau)}{\partial t}\right\rangle\,d\tau
\end{equation}
and inserting the equation of the phase fluctuations \eqref{eq:phase2}, we obtain 
\begin{equation}
\Delta\omega
=\frac{\alpha_\mathrm{H,eff}^2}{4\langle|f|^2\rangle^2}\int_{-\infty}^\infty \langle F_{|f|^2}(t)F_{|f|^2}(t+\tau)\rangle\,d\tau
+\int_{-\infty}^\infty\langle F_\varphi(t) F_\varphi(t+\tau)\rangle\,d\tau,
\end{equation}
where we have taken  \eqref{eq:f2phi} into account.
Inserting \eqref{eq:f2f2} and \eqref{eq:phiphi} results in the remarkably simple expression for the spectral linewidth
\begin{equation}
\label{eq:nu}
\Delta\nu=\frac{\Delta\omega}{2\pi}
=\frac{D_{f^* f}}{2\pi\langle|f|^2\rangle}\left(1+\alpha_\mathrm{H,eff}^2\right).
\end{equation}

In the following, we will rewrite this expression in terms of the intra-cavity photon number
\begin{equation}
\label{eq:photonnumber}
I_\mathrm{ph}=\frac{1}{\hbar\omega_0}\int_0^L P\,\Re{\left(\frac{\partial \beta}{\partial\omega}\right)} \,dz=\frac{P_0}{\hbar\omega_0}\frac{\int_0^L \Vert\Phi\Vert^2 \Re{\left(\frac{\partial \beta}{\partial\omega}\right)} \,dz}{(1-|r_0|^2)|\Phi^-(0)|^2}
\end{equation}
where 
\begin{equation}
P(z)=\langle|f|^2\rangle\Vert\Phi(z)\Vert^2
\end{equation}
is the stationary optical power inside the cavity and
\begin{equation} 
P_0=(1-|r_0|^2)\langle|f|^2\rangle|\Phi^-(0)|^2
\end{equation}
is the outcoupled power at $z=0$.
Moreover, we define the rate of spontaneous emission into the lasing mode
\begin{equation}
\label{eq:Rsp}
R_\mathrm{sp}=\frac{\int_0^L \Vert\Phi\Vert^2 n_\mathrm{sp}g\,dz}{\int_0^L \Vert\Phi\Vert^2 \Re{\left( \frac{\partial \beta}{\partial\omega} \right)} \,dz}
\end{equation}
and the longitudinal excess factor of spontaneous emission 
\begin{equation}
\label{eq:Petermann}
K=\frac{\left(\int_0^L \Vert\Phi\Vert^2\Re{\left(\frac{\partial \beta}{\partial\omega}\right)} \,dz\right)^2}{\left|(\Phi,\frac{\partial \beta}{\partial \omega}\Phi)\right|^2}.
\end{equation}
Now, \eqref{eq:nu} can be written as
\begin{equation}
\label{eq:nu2}
\Delta\nu=\frac{KR_\mathrm{sp}}{4\pi I_\mathrm{ph}}\left(1+\alpha_\mathrm{H,eff}^2\right)
\end{equation}
which is the standard form found in the literature \cite{tromborg1991theory,henry1986phase}.

The mode profile $\Phi(z)$ entering \eqref{eq:photonnumber}, \eqref{eq:Rsp}, and \eqref{eq:Petermann} is obtained by solving \eqref{eq:Phi}, \eqref{eq:threshold} and \eqref{eq:carrier_steady} self-consistently above threshold.
Thus, $\Phi$ is a mode of the active cavity affected by spatial hole burning.
The expression for the $K$--factor generalizes the one given in \cite{wenzel1996mechanisms} for non-uniform and complex--valued group velocity $(\partial \beta/\partial \omega)^{-1}$ and is in basic correspondence with \cite{tromborg1991theory,champagne1992global,wenzel1994equation}, but fundamentally different from the modified $K$--factor introduced in \cite{pick2015ab}.

At an exceptional point, the mode is orthogonal to itself, \textit{i.e.}, $(\Phi,\frac{\partial \beta}{\partial \omega}\Phi)=0$, and the $K$--factor approaches infinity. However, one has to keep in mind that at an exceptional point the single--mode approximation used here fails. Instead, for the description of the dynamics in the vicinity of such a point one has use at least one eigenmode and the corresponding generalized eigenmode \cite{wenzel1996mechanisms}.

\section{Impact of passive sections, chirp reduction factor and Fabry--P\'erot case}
\label{sec:Passiv}

In this section, we apply the general theory developed above to special configurations that allow for further simplifications (\textit{i.e.}, neglect of spatial hole burning, nonlinear gain and index as well gain dispersion). Thereby, it is shown that the general framework entails some well-established formulas from the literature as special cases.

\subsection{Linewidth of a laser consisting of a gain chip subject to feedback from an external cavity}

\begin{SCfigure}[][t]
\includegraphics[scale=0.95]{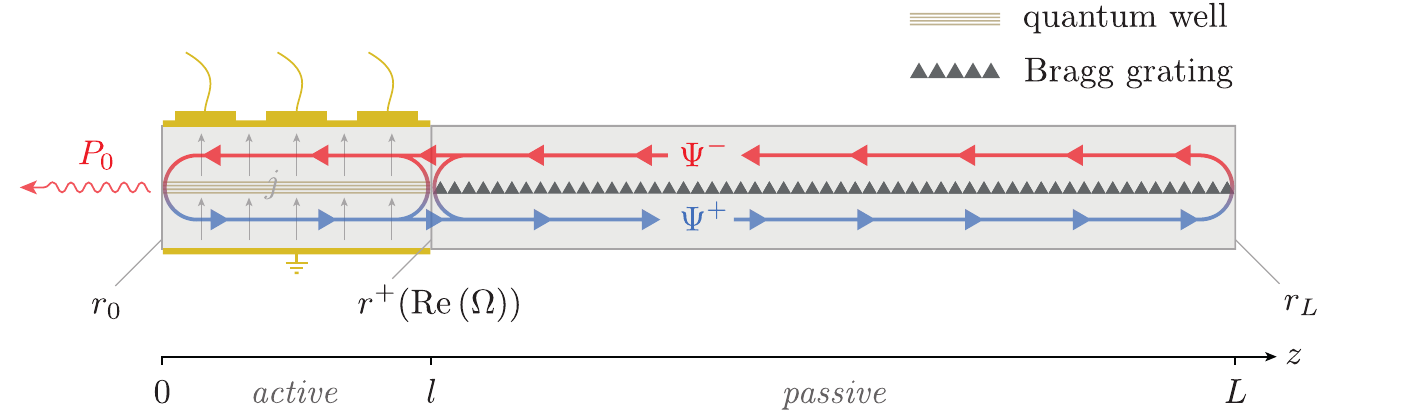}
\caption{Schematic illustration of a two-section DBR laser featuring an active gain section and a passive Bragg grating as considered in Sec.~\ref{sec:Passiv}.
\label{fig:DBRlaser}}
\end{SCfigure}

We consider a laser consisting of one active section for $z\in[0,l]$ with length $l$ and an arbitrary number of passive sections (including an external cavity) for  $z\in[l,L]$ with total length $L-l$.  The active section is of the Fabry--P\'erot (FP) type having no Bragg grating ($\kappa=0$ for $z\in[0,l])$. The passive sections may contain Bragg gratings but no excess carriers ($N=0$, $\partial \beta/\partial N=0$, $D_\mathrm{sp}=0$, $R_\mathrm{st}=0$ for $z\in[l,L]$). The reference plane is located just inside the active section so that a possible finite reflectivity at the interface between active and passive sections belongs to the passive sections.
Furthermore, we neglect spatial hole burning (\textit{i.e.}, $\partial N/\partial z=0$), nonlinear gain and index ($\partial \beta/\partial |f|^2=0$, $\partial R_\mathrm{st}/\partial |f|^2=R_\mathrm{st}/|f|^2$) and gain dispersion ($\beta_{\cal D}=0$, $\partial\beta/\partial\omega = n_\mathrm{g}/c$).
See Fig.~\ref{fig:DBRlaser} for an illustration of the device.

The functions $\Phi^\pm(z)$ as solution of \eqref{eq:Phi} are defined on the whole cavity $z\in[0,L]$. In the active section they are given by
\begin{equation}
\label{eq:modes_FP}
\Phi^\pm(z)|_\mathrm{active}=\Phi^\pm(l)\mathrm{e}^{\mp i(\Delta\overline{\beta}+\frac{n_{\mathrm{g}}}{c}\Omega)(z-l)}
\end{equation}
and hence 
\begin{equation}
\Phi^+(z)\Phi^-(z)|_\mathrm{active}=\Phi^+(l)\Phi^-(l)
\end{equation}
holds.
Furthermore we define the left and right reflectivities
\begin{equation}
\label{eq:rpm}
r^+=\frac{\Phi^-(l)}{\Phi^+(l)}\quad\mathrm{and}\quad r^-=\frac{\Phi^+(l)}{\Phi^-(l)}\equiv r_0\mathrm{e}^{-2il\left(\Delta\overline{\beta}+\frac{n_{\mathrm{g}}}{c}\Omega\right)},
\end{equation}
where we exploited the fact that the $\Phi^\pm(z)$ are continuous at $z=l$.
The derivative of $r^+$ can be expressed as (see Appendix \ref{appendix:proof})
\begin{equation}
\label{eq:lnrplus}
\frac{\partial\mathrm{ln}(r^+)}{\partial\omega}=-2i\frac{\int_{l}^L\Phi^+\Phi^-\frac{\partial \beta}{\partial \omega}\,dz}{\Phi^+(l)\Phi^-(l)}.
\end{equation}
Therefore,
\begin{equation}
\begin{aligned}
(\Phi,\frac{\partial \beta}{\partial \omega}\Phi)
&=2\int_0^l\Phi^+\Phi^-\frac{\partial \beta}{\partial \omega}\,dz+2\int_l^L\Phi^+\Phi^-\frac{\partial \beta}{\partial \omega}\,dz\\
&=2\Phi^+(l)\Phi^-(l)\left[l\frac{\partial \beta}{\partial \omega}\bigg|_\mathrm{active}+\frac{i}{2}\frac{\partial\mathrm{ln}(r^+)}{\partial\omega}\right]\\
&=2l\Phi^+(l)\Phi^-(l)\frac{\partial \beta}{\partial \omega}\bigg|_\mathrm{active}\chi
\end{aligned}
\end{equation}
where we introduced the parameter
\begin{equation}
\label{eq:chi}
\chi=1+\frac{i\frac{\partial\mathrm{ln}(r^+)}{\partial\omega}}{2l\frac{\partial \beta}{\partial \omega}}
=1+i\frac{c}{2ln_\mathrm{g}}\frac{\partial\mathrm{ln}(r^+)}{\partial\omega}
\end{equation}
as in \cite{tromborg1991theory}. We omit $|_\mathrm{active}$ in what follows.
Furthermore
\begin{equation}
\begin{aligned}
(\Phi,\tau_\mathrm{d}\frac{\partial \beta}{\partial N}\frac{\partial R_\mathrm{st}}{\partial |f|^2} \Phi)
& = \frac{2}{|f|^2}\int_0^l\Phi^+\Phi^-\tau_\mathrm{d}\frac{\partial \beta}{\partial N}R_\mathrm{st}\,dz\\
& = \frac{2}{|f|^2}\Phi^+(l)\Phi^-(l)\frac{\partial \beta}{\partial N}\int_0^l\tau_\mathrm{d}R_\mathrm{st}\,dz
\end{aligned}
\end{equation}
holds.
Then from \eqref{eq:halpha}, \eqref{eq:alphaHeff} and \eqref{eq:alfaH}
\begin{equation}
\label{eq:alphaHeff3}
\alpha_\mathrm{H,eff}
= -\frac{\mathrm{Re}\left(\frac{\frac{\partial\beta}{\partial N}}{\frac{\partial \beta}{\partial \omega}}\chi^\ast\right)}
{\mathrm{Im}\left(\frac{\frac{\partial\beta}{\partial N}}{\frac{\partial \beta}{\partial \omega}}\chi^\ast\right)}
= -\frac{\mathrm{Re}\left(\frac{\partial\beta}{\partial N}\chi^\ast\right)}
        {\mathrm{Im}\left(\frac{\partial\beta}{\partial N}\chi^\ast\right)}
= \frac{\mathrm{Re}\left(\alpha_\mathrm{H}\chi+i\chi\right)}
        {\mathrm{Im}\left(\alpha_\mathrm{H}\chi+i\chi\right)}		
\end{equation}
and
\begin{equation}
\label{eq:alphaeff2}
1+\alpha_\mathrm{H,eff}^2
= (1+ \alpha_\mathrm{H}^2 )\frac{|\chi|^2}{\mathrm{Im}^2\left(\alpha_\mathrm{H}\chi+i\chi\right)}
\end{equation}
follow.
Similarly, we obtain
\begin{equation}
\label{eq:Dff2}
\begin{aligned}
D_{f^\ast f}
&=\frac{\int_0^{l} \Vert\Phi\Vert^2 D_{\mathrm{sp}}\,dz}
{\left|2\int_0^{l}\Phi^+\Phi^-\frac{\partial \beta}{\partial \omega}\,dz\right|^2}
\frac{1}{\left|\chi\right|^2}.
\end{aligned}
\end{equation}
Hence, the product of \eqref{eq:alphaeff2} and \eqref{eq:Dff2} is given by
\begin{equation}
D_{f^\ast f}\left(1+\alpha_\mathrm{H,eff}^2\right)=\frac{\int_0^{l} \Vert\Phi\Vert^2 D_{\mathrm{sp}}\,dz}
{\left|2\int_0^{l}\Phi^+\Phi^-\frac{\partial \beta}{\partial \omega}\,dz\right|^2}
\frac{\left(1+\alpha_\mathrm{H}^2\right)}{\mathrm{Im}^2\left(\alpha_\mathrm{H}\chi+i\chi\right)}
\end{equation}
and the spectral linewidth can be written as
\begin{equation}
\label{eq:Dnupassiv}
\Delta\nu=\frac{\Delta\nu_\mathrm{FP}}{F^2}
\end{equation}
with the chirp reduction factor
\begin{equation}
\label{eq:F}
\begin{aligned}
F
&=\Im{\left(\alpha_\mathrm{H}\chi+i\chi\right)}\\
&=1-\frac{c}{2n_\mathrm{g}l}\Im{\left(\frac{\partial\mathrm{ln}(r^+)}{\partial\omega}\right)}+\alpha_\mathrm{H}\frac{c}{2n_\mathrm{g}l}\Re{\left(\frac{\partial\mathrm{ln}(r^+)}{\partial\omega}\right)}
\end{aligned}
\end{equation}
and 
\begin{equation}
\label{eq:DnuFP}
\Delta\nu_\mathrm{FP}=\frac{\int_0^{l} \Vert\Phi\Vert^2 D_{\mathrm{sp}}\,dz}
{2\pi\langle|f|^2\rangle\left|2\int_0^{l}\Phi^+\Phi^-\frac{\partial \beta}{\partial \omega}\,dz\right|^2}
\left(1+\alpha_\mathrm{H}^2\right)
\end{equation}
being the linewidth of a FP laser having the cavity length $l$ and the reflection coefficient $r^+(\Omega)$ at the rear facet.
Eq. \eqref{eq:F} agrees with \cite{komljenovic2015widely,tran2019tutorial} as well as \cite{kazarinov1987relation, boller2020hybrid} if the differently chosen harmonic time dependence ($\mathrm{e}^{i\omega_0 t}$ as used here versus $\mathrm{e}^{-i\omega_0 t}$) is observed. The relation to the static frequency chirp is given in 
Appendix {\ref{appendix:chirp_reduction}}.
Introducing an effective length of the passive cavity 
\begin{equation}
l_\mathrm{p}=-\frac{c}{2n_{\mathrm{g,p}}} \Im{\left(\frac{\partial\mathrm{ln}(r^+)}{\partial\omega}\right)},
\end{equation}
the chirp reduction factor can be written as
\begin{equation}
F=1+\frac{n_{\mathrm{g,p}} l_\mathrm{p}}{n_\mathrm{g} l}+\alpha_\mathrm{H}\frac{c}{2n_\mathrm{g}l}\Re{\left(\frac{\partial\mathrm{ln}(r^+)}{\partial\omega}\right)}.
\end{equation}
Thus the second term on the right--hand side of \eqref{eq:F} can be interpreted as the ratio of the round-trip times in the passive and active sections.
The chirp reduction factor allows an estimation of the reduction of the linewidth by a passive section or an external cavity for given group index, length and Henry's $\alpha$--factor of the active section.

\subsection{Linewidth of the Fabry--P\'erot laser cavity}

We start from \eqref{eq:nu2} and employ the same approximations resulting in \eqref{eq:Dnupassiv} and \eqref{eq:F}.
The rate of spontaneous emission into the lasing mode \eqref{eq:Rsp} is
\begin{equation}
\label{eq:RspFP}
R_\mathrm{sp}(N)=\frac{\int_0^l \Vert\Phi(z)\Vert^2 n_\mathrm{sp}(N)g(N)\,dz}{\int_0^l \Vert\Phi(z)\Vert^2 \Re{\left( \frac{\partial \beta}{\partial\omega}\right)} \,dz}
=\frac{c}{n_\mathrm{g}}n_{\mathrm{sp}}(N)g(N).
\end{equation}
The expression \eqref{eq:RspFP} should be compared with
\begin{equation}
\label{eq:Rsprate}
\tilde{R}_\mathrm{sp}(N)=\beta_\mathrm{sp}VR_\mathrm{rad}(N)
\end{equation}
often used in rate equation based modeling \cite{coldren2012}, where $\beta_\mathrm{sp}$ is the dimensionless spontaneous emission factor (ratio between the spontaneous emission going into the lasing mode and the spontaneous emission into all modes), $V$ the volume of the active region and $R_\mathrm{rad}=B N^2$ the rate of 
radiative spontaneous recombination ($B$ bimolecular recombination coefficient).
Equalizing \eqref{eq:RspFP} and \eqref{eq:Rsprate} at the lasing threshold yields
\begin{equation}
\beta_\mathrm{sp}=\frac{c}{n_\mathrm{g}}\frac{n_\mathrm{sp}(N_\mathrm{th})g(N_\mathrm{th})}{VR_\mathrm{rad}(N_\mathrm{th})},
\end{equation}
which is of the order $10^{-5}$ for typical edge--emitting lasers.

The photon number \eqref{eq:photonnumber} can be written as
\begin{equation}
\begin{aligned}
\label{eq:photonnumberFP1}
I_\mathrm{ph}
&=\frac{P_0}{\hbar\omega_0}\frac{\int_0^l \Vert\Phi(z)\Vert^2 \Re{\left( \frac{\partial \beta}{\partial\omega} \right)}\,dz}{(1-|r_0|^2)|\Phi^-(0)|^2}\\
&=\frac{n_\mathrm{g}}{c}\frac{P_0}{\hbar\omega_0\left( (1-|r_0|^2)|\Phi^-(0)|^2 \right) }\int_0^l\left[ |\Phi^+(0)|^2\mathrm{e}^{2\Im{(\overline{\beta})}z}
+|\Phi^-(0)|^2\mathrm{e}^{-2\Im{(\overline{\beta})}z} \right]\,dz\\
&=\frac{n_\mathrm{g}}{c}\frac{P_0}{\hbar\omega_0}
\frac{|r_0|^2\left(\mathrm{e}^{2\Im{(\overline{\beta})}l}-1\right)
-\mathrm{e}^{-2\Im{(\overline{\beta})}l}+1}{2\Im{(\overline{\beta})}(1-|r_0|^2)}.
\end{aligned}
\end{equation}
The complex mode frequencies of the FP cavity are the solutions \footnote{They are obtained from \eqref{eq:modes_FP} applying the boundary conditions \eqref{eq:bc2} at $z=0$ and $z=l$.} of 
\begin{equation}
\label{eq:cmplxfreqFP}
\Omega_m=\frac{c}{n_\mathrm{g}}\left[\frac{m\pi}{l}-\overline{\beta}-\frac{i}{2l}\ln{\left(r_0r^+(\Omega_m) \right)}\right]
\end{equation}
with $m\in\mathbb{Z}$. 
To ensure $\Im{(\Omega)}=0$ for the lasing mode, \eqref{eq:cmplxfreqFP} leads to the usual threshold condition
\begin{equation}
\label{eq:thresholdFP}
\alpha_\mathrm{out}\equiv-\frac{1}{l}\ln(|r_0r^+(\Re{(\Omega)})|)=2\Im{(\overline{\beta})}\equiv g-\alpha.
\end{equation}
In what follows, we set $r_l=r^+(\Omega)$ (reflectivity seen by the laser at $z=l$).
From \eqref{eq:thresholdFP} it follows
\begin{equation}
\label{eq:photonnumberFP2}
I_\mathrm{ph}
=\frac{n_\mathrm{g}}{c}\frac{P_0}{\hbar\omega_0}
\frac{|r_0|^2\left(\frac{1}{|r_0r_l|}-1\right)
-|r_0r_l|+1 }{2\Im{(\overline{\beta})}(1-|r_0|^2)}
=\frac{n_\mathrm{g}}{c}\frac{P_\mathrm{out}}{\hbar\omega_0\alpha_\mathrm{out}},
\end{equation}
where we used the relation between outcoupled power $P_0$ and total output power $P_\mathrm{out}$ 
\begin{equation}
\label{eq:P0}
P_0=\frac{P_\mathrm{out}}{1+\frac{1-|r_l|^2}{1-|r_0|^2}\frac{|r_0|}{|r_l|}}.
\end{equation}
The final result is
\begin{equation}
\label{eq:Dnu2FP2}
\Delta\nu_\mathrm{FP}=\frac{K_\mathrm{FP}}{4\pi }\frac{c^2}{n_\mathrm{g}^2}\frac{\hbar\omega_0n_{\mathrm{sp}}
(\alpha+\alpha_\mathrm{out})\alpha_\mathrm{out}}{P_\mathrm{out}}\left(1+\alpha_\mathrm{H}^2\right),
\end{equation}
where Petermann's $K$--factor 
\begin{equation}
K=
\frac{\left(\int_0^l \Vert\Phi\Vert^2 \Re{\left(\frac{\partial \beta}{\partial\omega}\right)} \,dz\right)^2}{\left|2\int_0^l\Phi^+\Phi^-\frac{\partial \beta}{\partial \omega}\,dz\right|^2}
\end{equation}
of the FP cavity can be similarly derived as
\begin{equation}
K_\mathrm{FP}
=\left[\frac{(|r_0|+|r_l|)(1-|r_0r_l|)}
{2|r_0r_l|\ln(|r_0r_l|)}
\right]^2
\end{equation}
in agreement with \cite{henry1986theory, hamel1989nonorthogonality}.
Eq. \eqref{eq:Dnu2FP2} together with \eqref{eq:P0} was used by several authors, \textit{e.g.}, \cite{henry1982theory, spiessberger2011dbr, boller2020hybrid}.
However, one should be aware of the approximations involved in the derivation.

The original Schawlow--Townes formula \cite{schawlow1958infrared} is obtained by setting
$K_\mathrm{FP}=1$, $n_{\mathrm{sp}}=1$, $\alpha=0$, $\alpha_\mathrm{H}=0$, and relating the 
outcoupling losses $\alpha_\mathrm{out}$ to the spectral FWHM of a cavity resonance by 
$\Delta\nu_\mathrm{cav}=c/n_\mathrm{g}\cdot\alpha_\mathrm{out}/2\pi$ to
\begin{equation}
\label{eq:nuST}
\Delta\nu_\mathrm{ST}=\frac{\pi\hbar\omega_0}{P_\mathrm{out}}(\Delta\nu_\mathrm{cav})^2.
\end{equation}
There are two reasons why this differs from the original formula by a factor of $4$.
First, in \cite{schawlow1958infrared} the half widths instead of full widths are used (factor of $2$).
Second, in \cite{schawlow1958infrared} the sub-threshold case is considered resulting in a further factor of $2$,
which can be seen as follows: Below threshold the spectral density can be directly calculated from \eqref{eq:fnew} for $\Delta M=0$ by Fourier transformation via
\begin{equation}
\langle f^\ast(\omega)f(\omega')\rangle=2\pi S_{f}(\omega) \delta(\omega-\omega')
\end{equation}
with the result
\begin{equation}
\label{eq:Sf}
S_f(\omega)=\frac{2D_{f^* f}}{\left(\omega-\Re{(\Omega)} \right)^2+\left(\Im{(\Omega)}\right)^2 }
=\frac{2D_{f^* f}}{(\omega-\Re{(\Omega)} )^2+\left(\frac{D_{f^* f}}{\langle|f|^2\rangle} \right)^2},
\end{equation}
where we have used \eqref{eq:mean} and \eqref{eq:Fff}.
This is a Lorentzian with the FWHM
\begin{equation}
\label{eq:Dnubelow}
\Delta\nu_\mathrm{subthr}=\frac{D_{f^* f}}{\pi\langle|f|^2\rangle},
\end{equation}
which is a factor of $2$ larger than \eqref{eq:nu} (for $\alpha_\mathrm{H,eff}=0$). See \cite{risken1968statistik} for a more thorough discussion.

\begin{figure}[t]
\centering
\includegraphics[scale=0.85]{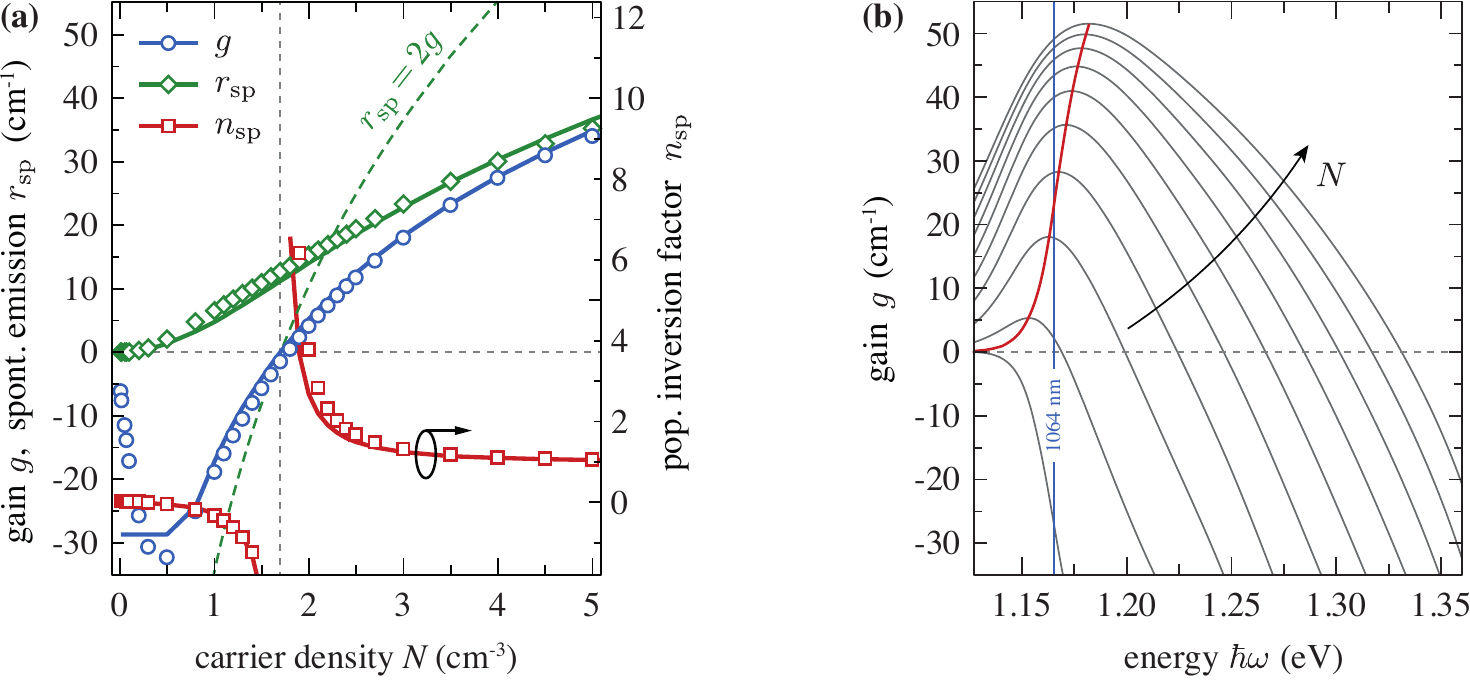}
\caption{\textbf{(a)}~Modal gain (blue) and rate of spontaneous emission (green), left axis, and population inversion factor (red), right axis, versus carrier density obtained from a microscopic simulation (symbols) and the analytic models \eqref{eq:gainmodel}, \eqref{eq:rspmodel} and $n_\mathrm{sp}=r_\mathrm{sp}/g$ (solid lines) for a fixed wavelength $\lambda_0=1064\,\text{nm}$.
The dashed green line is the spontaneous emission computed using a constant population inversion factor $n_\mathrm{sp}=2$. See the Appendix \ref{appendix:dispersion} for the consideration of the gain at a fixed wavelength within the framework of the present model. \textbf{(b)}~Microscopically computed gain spectra for carrier densities ranging from $N=1\cdot 10^{18}\,\text{cm}^{-3}$ to $1\cdot 10^{19}\,\text{cm}^{-3}$. The position of the fixed wavelength $\lambda_0$ is shown by a blue line, the peak gain position is indicated by a red line.
\label{fig:gain_rsp_nsp}}
\end{figure}  

\section{Population inversion factor}
\label{sec:spont}

The population inversion factor\footnote{Sometimes it is called \lq spontaneous emission factor\rq\ which could be misleading.} introduced in \eqref{eq:Dsp} and used in \eqref{eq:Rsp} and \eqref{eq:Dnu2FP2} is given by \cite{lasher1964spontaneous}
\begin{equation}
n_\mathrm{sp}(N)=\frac{1}{1-\exp{\left(\frac{\hbar\omega_0-q U_\mathrm{F}(N)}{k_\mathrm{B}T} \right)}},
\end{equation}
where $U_\mathrm{F}$ is the Fermi voltage (spacing of quasi--Fermi potentials of holes and electrons), $q$ the elementary charge, $k_\mathrm{B}$ the Boltzmann constant and $T$ the temperature.
It has a singularity at the transparency density $N_\mathrm{tr}$, where $\hbar\omega_0=U_\mathrm{F}(N_\mathrm{tr})$ and $g(N_\mathrm{tr})=0$. 
Therefore, the replacement
\begin{equation}
r_\mathrm{sp}(N)\equiv n_\mathrm{sp}(N)g(N)=n_\mathrm{sp}(\alpha+\alpha_\mathrm{out})
\end{equation}
with constant $n_\mathrm{sp}$ often employed is critical because for high--$Q$--cavities with low $\alpha$ and $\alpha_\mathrm{out}$
the spontaneous emission is underestimated.
We calculated the modal gain, the spontaneous emission into the lasing mode and the inversion factor employing a 
numerical simulation based on a $8\times8$ $\mathbf{k\cdot p}$ band structure calculation and a free carrier theory for the optical response functions with phenomenological corrections for many-body effects such as band--gap renormalization, Coulomb enhancement and transition broadening \cite{wenzel1999improved}.
The results for a 5 nm thick InGaAs quantum well emitting around $1064~$nm embedded into an AlGaAs-based waveguide structure are compared with analytical models in Fig.~\ref{fig:gain_rsp_nsp}.
The increase of the absorption (negative gain) at small increasing carrier densities observed in the simulation is caused by band--gap narrowing, shifting the absorption edge to longer wavelengths \cite{wenzel2000effect}.

The gain at a fixed wavelength is modeled as 
\begin{equation}
\label{eq:gainmodel}
g=g^\prime N_\mathrm{tr}\ln{\left[\frac{\mathrm{max}(N,N_\mathrm{cl})}{N_\mathrm{tr}}\right]}
\end{equation}
and the modal spontaneous emission as
\begin{equation}
\label{eq:rspmodel}
r_\mathrm{sp}=\frac{g^\prime N_\mathrm{tr}}{2}\ln{\left[1+\left(\frac{N}{N_\mathrm{tr}}\right)^2\right]}
\end{equation}
with differential gain $g^\prime=19\cdot10^{-22}~\mathrm{m}^2$, transparency density $N_\mathrm{tr}=1.7\cdot10^{24}~\mathrm{m}^{-3}$ and gain clamping density $N_\mathrm{cl}=7\cdot10^{23}~\mathrm{m}^{-3}$.
Eq.~\eqref{eq:rspmodel} exhibits the correct asymptotic behavior $r_\mathrm{sp}\propto N^2$ for $N\ll N_\mathrm{tr}$ and $r_\mathrm{sp} = g$ for $N\to\infty$.
The agreement of $r_\mathrm{sp}$ between simulation and model is remarkably good without the need of introducing any new parameters.
The singularity of the inversion factor at $N=N_\mathrm{tr}$ is clearly visible in Fig.~\ref{fig:gain_rsp_nsp}.
If a constant value of $n_\mathrm{sp}=2$ is used, the spontaneous emission is considerably under-- or over--estimated (depending on the respective carrier density).

\section{Numerical results for a DBR laser} \label{sec:example}

\begin{table}[t] 
\begin{tabular*}{0.667\linewidth}{@{}c c c c@{\extracolsep{\fill}}c@{}}
\toprule
\textbf{parameter}	& \textbf{symbol}	& \textbf{unit} &
\textbf{value} & \textbf{first used}\\
\midrule
reference wavelength	& $\lambda_0$	& m  &  $1.064\cdot10^{-6}$ &  \\
front facet reflectivity	& $|r_0|^2$ 
&    &  $0.3$ & \eqref{eq:bc} \\
rear facet reflectivity	  & $|r_L|^2$ 
	&    &  $0$ & \eqref{eq:bc} \\
internal optical loss (both sections)  & $\alpha$& m$^{-1}$ & $60$ & \eqref{eq:betaN} \\
group index (both sections)           & $n_\mathrm{g}$  &  & 3.9 & \eqref{eq:cwe} \\
built-in index detuning (both sections)          & $\Delta n_0$  &  & 0 & \eqref{eq:ThermTuning} \\
\midrule
\multicolumn{5}{c}{\emph{active section}}\\
\midrule
length                  	& $l$			& m  &  $1\cdot10^{-3}$ &  \\
differential gain	        & $g'$   & m$^{2}$ & $18.62\cdot10^{-22}$ & \eqref{eq:gainmodel} \\
$\alpha$--factor          & $\tilde{\alpha}_\mathrm{H}$&  & $1$ &  \eqref{eq:alfaH}\\
transparency carrier density  & $N_\mathrm{tr}$   & m$^{-3}$ & $1.7\cdot10^{24}$ & \eqref{eq:gainmodel} \\
self-heating induced index tuning           & $\nu_s$  &
A$^{-1}$ & $9.16 \cdot 10^{-3}$  & \eqref{eq:ThermTuning} \\
gain clamping density  & $N_\mathrm{cl}$   & m$^{-3}$ & $0.8\cdot10^{24}$ & \eqref{eq:gainmodel}  \\
index clamping density  & $N_\mathrm{cl,i}$   & m$^{-3}$ & $0.01\cdot10^{24}$ & \eqref{eq:IndexModel} \\
gain saturation power  & $P_\mathrm{sat}$ & W & $5.866$ & \eqref{eq:GainModelSat} \\
dispersion peak amplitude  & $g_{\cal D}$ & m$^{-1}$ & $50$ & \eqref{eq:GainDisp} \\
dispersion peak frequency detuning  & $\omega_{\cal D}$ &rad\,s$^{-1}$ & $0$ & \eqref{eq:GainDisp} \\
dispersion HWHM  & $\gamma_{\cal D}$ &rad\,s$^{-1}$ & $83.25\cdot 10^{12}$ & \eqref{eq:GainDisp} \\
thickness of active region  & $d$   & m & $4\cdot10^{-6}$ & \eqref{eq:carrier} \\
width of active region  & $W$   & m & $10\cdot10^{-9}$ &  \eqref{eq:rstim_orig}\\
series resistance  & $R_s$   & $\Omega$ & $1$ & \eqref{eq:currentdensity} \\
Fermi voltage derivative & $\frac{dU_\mathrm{F}}{dN}$   & Vm$^3$ & $3\cdot 10^{-26}$ & \eqref{eq:currentdensity2} \\
defect recombination coefficient & $A$   & s$^{-1}$ & $4\cdot 10^{8}$ & \eqref{eq:SpontRec} \\
bimolecular recombination coefficient & $B$   & m$^3$\,s$^{-1}$  & $1\cdot 10^{-16}$& \eqref{eq:SpontRec} \\
Auger recombination coefficient  & $C$    & m$^6$\,s$^{-1}$ & $4\cdot 10^{-42}$& \eqref{eq:SpontRec} \\
injection current  & $I$   & A & $[0,300]\cdot 10^{-3}$ & \eqref{eq:currentdensity} \\
\midrule
\multicolumn{5}{c}{\emph{passive section}}\\
\midrule
length                  	& $L-l$			& m  &  $3\cdot10^{-3}$ &  \\
coupling coefficient      & $\kappa$		& m$^{-1}$ & $200$ & \eqref{eq:matrixMbar} \\
cross-heating induced index tuning   & $\nu_c$  & A$^{-1}$ & $1.83 \cdot 10^{-3}$  & \eqref{eq:ThermTuning} \\
\bottomrule
\end{tabular*}
\caption{Parameters of the simulated DBR laser. 
\label{tab:DBR}}
\end{table}

In the simulations described below we study a simple two-section DBR laser as sketched in Figs. \ref{fig:sketch} and \ref{fig:DBRlaser}.
It consists of a $1$~mm long active gain section (without Bragg grating) and a $3$~mm long passive Bragg reflector section (without active layer) with a coupling coefficient of $\kappa=2~\text{cm}^{-1}$. 
The usage of such DBR lasers having long reflector sections with low coupling coefficients was suggested in Ref.~\cite{kazarinov1987relation} for the first time. 
In fact, in Ref.~\cite{spiessberger2011dbr} an intrinsic linewidth of $2$~kHz at an output power of $180$~mW was reported. This device had a $3$~mm long gain section and a $1$~mm long reflector section, where the active layer extended over the whole cavity (\textit{i.e.}, including the reflector section).
The device studied here resembles the one presented in Ref.~\cite{brox2021}, where an intrinsic linewidth of 
$4$~kHz at an output power of $73$~mW was reported. Experimentally, the intrinsic linewidth is determined from the plateau of the frequency noise spectrum, as in the theory \cite{schiemangk2014accurate}.

Within the active section, we assume the logarithmic gain model \eqref{eq:gainmodel} 
supplemented with the nonlinear gain saturation factor
\begin{equation}
\label{eq:GainModelSat}
g(z,N,\Vert\Psi\Vert^2) = g^\prime(z) N_\mathrm{tr}(z) \ln{\left[\frac{\mathrm{max}(N,N_\mathrm{cl}(z))}{N_\mathrm{tr}(z)}\right]}
\frac{1}{1+\Vert\Psi\Vert^2/P_\mathrm{sat}(z)} ,
\end{equation}
and
the refractive index model with square--root--like dependency on the carrier density
\begin{equation}
\label{eq:IndexModel}
\Delta n_N = 
-\frac{\lambda_0\tilde{\alpha}_\mathrm{H} g^\prime N_\mathrm{tr}}{2\pi} 
\left( \sqrt{\frac{\max(N,N_\mathrm{cl,i})}{N_\mathrm{tr}}} - 1\right).
\end{equation}
Note, that at the transparency carrier density $N_\mathrm{tr}$,
the fixed parameter $\tilde{\alpha}_\mathrm{H}$ agrees with the $\alpha$--factor $\alpha_\mathrm{H}$ from \eqref{eq:alfaH}, whereas  $\Delta n_N$ vanishes.
The $\alpha$--factor correspondingly increases for $N>N_\mathrm{tr}$. 
The value of $\tilde{\alpha}_\mathrm{H}=1$ was chosen in agreement with simulations \cite{wenzel1999improved} and is in basic agreement with measurements, where $\alpha$--factors between $1$ and $2$ for a highly-strained InGaAs quantum well were obtained, depending on the detuning \cite{batrak2005simulation}.

\begin{figure}[t]
\centering
\includegraphics[scale=0.89]{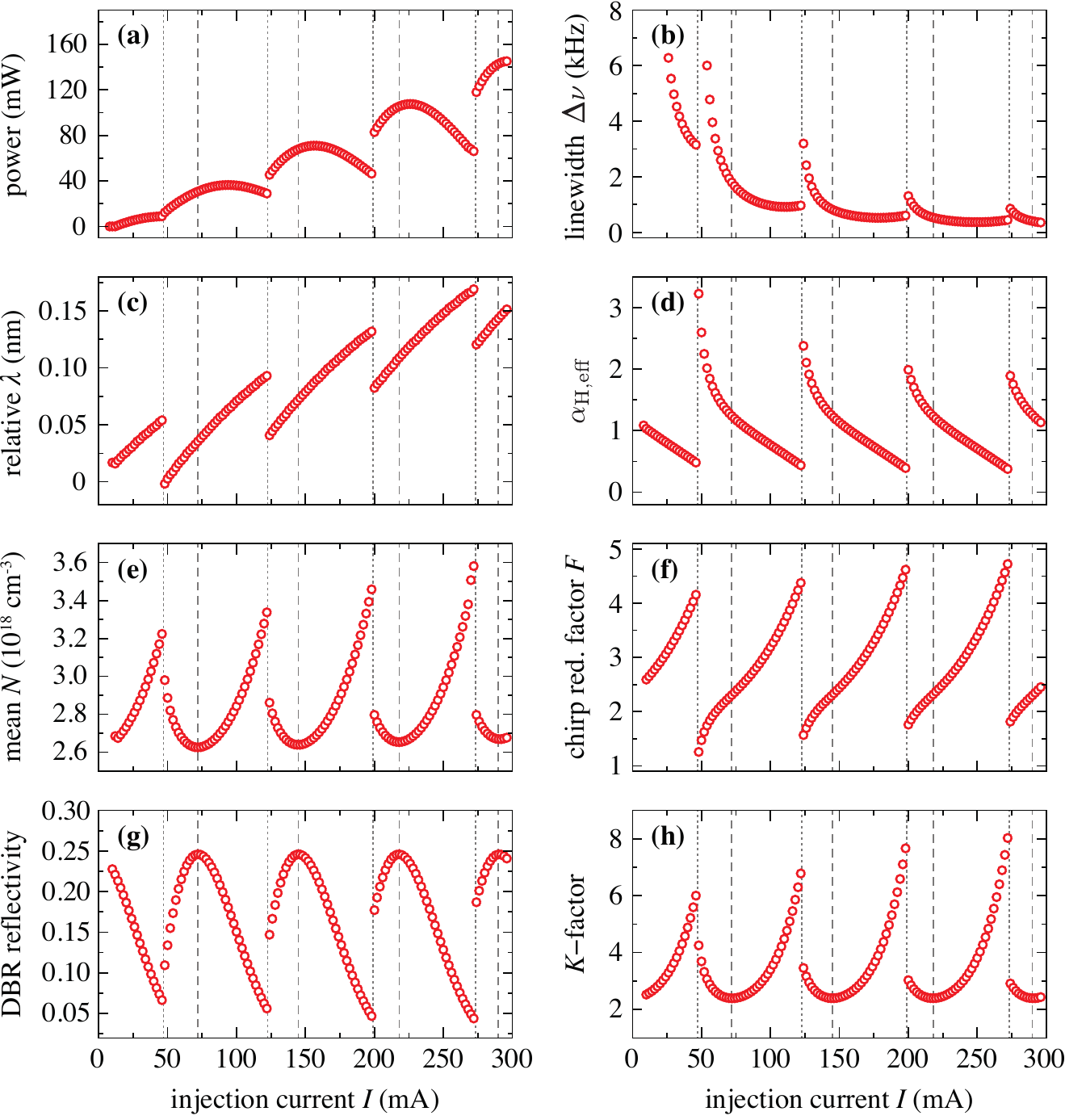}
\caption{Simulated characteristics of the DBR laser as functions of the upsweeped injection current.
Dotted and dashed lines indicate mode jumps and  maxima of the DBR reflectivity, respectively.
\label{fig:testDBR}}
\end{figure}   

Other changes of the refractive index, including 
self--heating--induced self- and cross--heating contributions,
are modeled by \cite{radziunas2017traveling}
\begin{equation}
\label{eq:ThermTuning}
\Delta n(z) = \Delta n_0(z)+ \Delta n_T(z),\qquad
\Delta n_T(z)=\begin{cases}
\nu_sI,  &z\in[0,l]
\cr  
\nu_cI,  &z\in[l,L]
\end{cases},
\end{equation}
where $I$ is the injection current into the active (gain) section. 
We employ the commonly used cubic model for the rate of non-radiative and spontaneous recombination of the carriers
in the active section
\begin{equation}
\label{eq:SpontRec}
R(N) = AN+BN^2+CN^3,
\end{equation} is approximated by
describing Shockley--Read--Hall recombination, direct band-to-band recombination and Auger recombination.
Finally, the current density \eqref{eq:currentdensity} in the gain section is approximated by 
\begin{equation}
\label{eq:currentdensity2}
j(z,N)=\frac{I}{Wl}-\frac{dU_\mathrm{F}}{dN}\;\frac{N-\frac{1}{l}\int_{0}^lN\,dz}{WlR_s},
\end{equation}
where the derivative of the Fermi voltage $dU_\mathrm{F}/dN$ is taken at a fixed carrier density.

The numerical solution of the coupled--wave equations
\eqref{eq:cwe}, \eqref{eq:bc}, \eqref{eq:carrier}, \eqref{eq:PolarizationEqs}
in the time domain were performed with the software package \texttt{LDSL-tool}
developed at the Weierstrass Institute \cite{LDSL}. The set of parameters
characterizing the simulated DBR laser is listed in Table~\ref{tab:DBR}.

Several characteristics of the steady
states obtained in the numerical simulations of the DBR laser with an upsweep of bias
current are shown in Fig.~\ref{fig:testDBR}. The almost-periodic jumps towards the steady-state defined by the
adjacent longitudinal optical mode and corresponding changes of the state
characteristics are induced by the self--and cross--heating modeled according to
Eq.~\eqref{eq:ThermTuning}.
While the self-heating of the active section is mainly responsible for the
fast shift of the lasing wavelength to longer values and periodic jumps to the
shorter-wavelength state, the cross--heating induces a reduced shift of the peak wavelength of the
DBR reflectivity and, therefore, the corresponding shift of the mean
lasing wavelength, see Fig.~\ref{fig:testDBR}\,(c).
The maximum emitted optical power, see Fig.~\ref{fig:testDBR}\,(a), and the minimal (averaged) carrier density, see Fig.~\ref{fig:testDBR}\,(e),
within each period coincide well with the maximum field reflection
provided by the Bragg grating shown in Fig.~\ref{fig:testDBR}\,(g).

The overall decay of the estimated spectral linewidth, see Fig.~\ref{fig:testDBR}\,(b), reflects its inverse
proportionality to the field intensity, see Eq.~\eqref{eq:nu}. The decay of the
linewidth within each period is consistent with the decrease of the modulus of the
effective linewidth enhancement factor $\alpha_{\mathrm{H},\mathrm{eff}}$, see Fig.~\ref{fig:testDBR}\,(d), the increase of the chirp reduction factor
$F$ shown in Fig.~\ref{fig:testDBR}\,(f), and the behavior of Petermann's $K$--factor  plotted in Fig.~\ref{fig:testDBR}\,(h).
The narrowest linewidth within each period is observed at the long--wavelength flank of the DBR reflectivity just before the transition to the neighboring state.
The dependence of $F$ and $\alpha_{\mathrm{H}}$ on the deviation of the lasing wavelength from the peak wavelength of the DBR reflectivity confirms earlier findings \cite{kazarinov1987relation, patzak1985analysis, chacinski2010impact} that $F$ increases and $\alpha_{\mathrm{H,eff}}$ decreases at the long--wavelength flank of the reflection spectrum.

\section{Outlook}\label{sec:outlook}

The theory presented can be easily extended to take into account carrier noise. 
Including the source $F_N \ne 0$, we obtain instead of \eqref{eq:deltaN} the relation
\begin{equation}
\label{eq:deltaN3}
\delta N= \tau_\mathrm{d}F_N - \tau_\mathrm{d}\frac{\partial R_\mathrm{st}}{\partial |f|^2}\delta|f|^2.
\end{equation}
Substituting this expression into \eqref{eq:fbetrag} yields a simple modification of the Langevin force
\begin{equation}
\label{eq:FNeff}
F_{|f|^2}\to F_{|f|^2}+2\mathrm{Im}\frac{(\Phi,\frac{\partial \beta}{\partial N}\tau_\mathrm{d}F_N\Phi)}
{(\Phi,\frac{\partial \beta}{\partial \omega}\Phi)}\langle|f|^2\rangle,
\end{equation}
leading to two additional additive contributions to the linewidth due to carrier noise and the cross-correlation between carrier and intensity noise.
The same procedure can be applied to the calculation of the modulation of amplitude and frequency in response to an external modulation of the current injection term in \eqref{eq:carrier}.

In case of gain coupling, \eqref{eq:DeltaM} has to be replaced by
\begin{equation}
\Delta M=\begin{bmatrix}
\overline{\beta}-\beta  &
\overline{\kappa}-\kappa \\
\overline{\kappa}-\kappa &
\overline{\beta}-\beta
\end{bmatrix}
\end{equation}
and \eqref{eq:variationbeta} has to be supplemented by
\begin{equation}
\kappa=\kappa(\langle N\rangle)+\frac{\partial \kappa}{\partial N}\delta N.
\end{equation}
Note that for loss or gain coupling or for higher order gratings there are additional contributions to the imaginary part of $\beta$.
Furthermore, the rate of stimulated recombination \eqref{eq:rstim_orig} has to be modified \cite{jonsson1996instabilities} and the diffusion coefficient \eqref{eq:Dff} is varied because of non--vanishing  $\langle F_\mathrm{sp}^{-\ast}(z,t) F_\mathrm{sp}^{+}(z',t') \rangle \neq 0$ and its complex conjugate \cite{baets1993distinctive,tromborg1994traveling}.

Fluctuations of the shape of the power profile can be accounted for by a linearization of \eqref{eq:cwe} and \eqref{eq:carrier} around a steady state, performing a Fourier transformation and solving for the fluctuations $\delta\Psi$, \textit{e.g.}, by the Green's function method \cite{tromborg1994traveling} to calculate the noise spectral densities $S_{\xi}(\tilde{\omega})$ from
\begin{equation}
\label{eq:noise_spectrum}
\langle \xi^\ast(\tilde{\omega})\xi(\tilde{\omega}')\rangle=2\pi S_{\xi}(\tilde{\omega}) \delta(\tilde{\omega}-\tilde{\omega}'),
\end{equation}
where $\xi$ is the variable of interest (\textit{e.g.}, phase or intensity).

\section{Summary}

We have derived in a self-contained manner the spectral linewidth of edge--emitting multi--section semiconductor lasers starting from the time-dependent coupled--wave equations with a Langevin noise source. For this, we have expanded the forward and backward propagating fields into the longitudinal modes of the open cavity and have obtained very general expressions for the effective linewidth enhancement factor $\alpha_{\mathrm{H}}$ and the longitudinal excess factor of spontaneous emission ($K$--factor), including the effects of nonlinear gain (gain compression) and index (Kerr effect), gain dispersion, longitudinal spatial hole burning for multi--section cavity structures.
We have shown that the general linewidth expression contains the effect of linewidth narrowing due to an external cavity by the chirp reduction factor $F$. 
Finally, we have investigated the dependence of the population inversion factor on the carrier density and proposed a new analytical formula.  
Based on the derived expressions, the spectral linewidth as well as $\alpha_{\mathrm{H}}$, $K$, and $F$ have been calculated for a two--section DBR laser as functions of the injection current and the optical output power, taking into account the thermal detuning between gain and reflector sections. The mode jumps appearing with increasing current are accompanied by sudden rises of the spectral linewidth due to the rise of the modulus of $\alpha_{\mathrm{H}}$ and the drop of $F$.

\acknowledgments{This research was partially funded by the Federal Ministry of Education and Research (BMBF) under grant number 50RK1972 and by the Deutsche Forschungsgemeinschaft (DFG, German Research Foundation) under Germany's Excellence Strategy – The Berlin Mathematics Research Center MATH+ (EXC-2046/1, project ID: 390685689, grant AA2-13).
The authors are indebted to Hans-J\"{u}rgen W\"{u}nsche for a critical reading of the manuscript.}

\appendix

\section{Dispersion operator}
\label{appendix:dispersion}

The dispersion operator $\widehat{\mathcal{D}}(z,t)$ in Eq.~\eqref{eq:cwe} models the
frequency dependence of the gain and associated refractive index.
The approach followed here is based on approximating the material gain spectrum locally around the gain peak by a Lorentzian and supplementing the time-domain coupled--wave equations \eqref{eq:cwe} with additional dynamical equations for \emph{auxiliary} polarizations \cite{ning1997effective, bandelow2001impact, sieber1998travelling}
\begin{equation}
\label{eq:PolarizationEqs}
\frac{\partial}{\partial t}P(z,t) = \left( i\omega_{\cal D}(z) - \gamma_{\cal D}(z)\right) P(z,t) + \gamma_{\cal D}(z)\Psi(z,t),
\end{equation}
where $\omega_{\mathcal{D}}$ is the detuning between the peak gain frequency $\omega_p$ and the reference frequency $\omega_0$, $\gamma_{\mathcal{D}}$ is the half width at half maximum (HWHM) of the gain curve and
\begin{equation*}
P(z,t) =
\begin{bmatrix}
P^+(z,t) \\
P^-(z,t)
\end{bmatrix}.
\end{equation*}
The corresponding dispersion operator in \eqref{eq:cwe} reads
\begin{equation}
\label{eq:DisperionOperator}
\widehat{\cal D}(z,t)\Psi(z,t) = -\frac{i}{2} g_{\cal D}(z) \left( \Psi(z,t) - P(z,t)\right),
\end{equation}
where $g_{\cal D}$ describes the difference between the maximum amplitude gain $g(z,N,\Vert \Psi \Vert^2)$, see Eq.~\eqref{eq:betaN}, and the gain at the reference frequency.
The parameters $\omega_{\mathcal{D}}$, $\gamma_{\mathcal{D}}$, and $g_{\mathcal{D}}$ are obtained from a fit to a microscopically computed gain spectrum or experimental data.
\begin{SCfigure}[][t]
\includegraphics[scale=0.95]{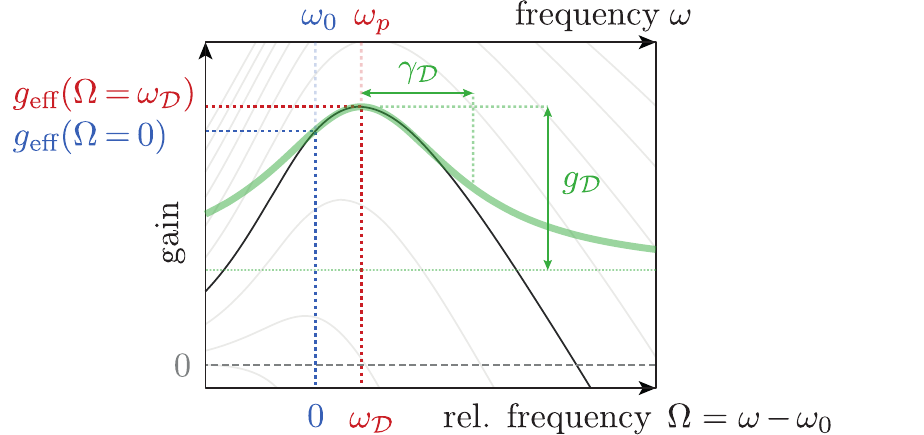}
\caption{Illustration of the Lorentzian fitted (effective) gain spectrum and labeling of the dispersion parameters $\omega_{\mathcal{D}}$, $\gamma_{\mathcal{D}}$, and $g_{\mathcal{D}}$. The peak gain is denoted by 
$g_{\text{eff}}\left(\Omega=\omega_{\mathcal{D}}\right)=g$ and $g_{\text{eff}}\left(\Omega=0\right)$ is the effective gain at the reference frequency $\omega_0$. \label{fig:dispersion}\vspace*{2.5ex}}
\end{SCfigure}   
A Fourier transformation of \eqref{eq:DisperionOperator} yields, together with the frequency domain solution of the polarization equation \eqref{eq:PolarizationEqs}, an expression for the dispersion factor (complex Lorentzian)
\begin{equation}
\label{eq:GainDisp}
\beta_{\mathcal{D}}\left(z,\Omega\right) = \frac{g_{\mathcal{D}}(z)}{2}\frac{\Omega-\omega_{\mathcal{D}}\left(z\right)}{i\left(\Omega-\omega_{\mathcal{D}}\left(z\right)\right)+\gamma_{\mathcal{D}}\left(z\right)}.
\end{equation}
The associated \emph{effective gain spectrum} entering the rate of stimulated emission \eqref{eq:rstim}
reads
\begin{equation}
\label{eq:geff}
\begin{aligned}
g_{\text{eff}}(z,N,\Vert\Psi\Vert^{2},\Omega) &= g(z,N,\Vert\Psi\Vert^{2}) + 2\Im{\left(\beta_{\mathcal{D}}(z,\Omega)\right)}\\
&=g(z,N,\Vert\Psi\Vert^{2}) - g_{\mathcal{D}}\frac{\left(\Omega-\omega_{\mathcal{D}}\left(z\right)\right)^{2}}{\left(\Omega-\omega_{\mathcal{D}}\left(z\right)\right)^{2}+\gamma_{\mathcal{D}}^{2}\left(z\right)},
\end{aligned}
\end{equation}
which is maximal at $\Omega=\omega_{\mathcal{D}}$, \textit{i.e.}, if the laser operates at the peak gain frequency.

An alternative representation of Eq.~\eqref{eq:DisperionOperator} is obtained with the help of the Green's function of the polarization equations \eqref{eq:PolarizationEqs} as
\begin{equation}
\label{eq:convolution}
\widehat{\mathcal{D}}\left(z,t\right)\Psi\left(z,t\right)=-\frac{i}{2}g_{\mathcal{D}}(z)\left(\Psi\left(z,t\right)-\gamma_{\mathcal{D}}(z)\int_{0}^{t}\mathrm{e}^{\left(i\omega_{\mathcal{D}}(z)-\gamma_{\mathcal{D}}(z)\right)\tau}\Psi\left(z,t-\tau\right) \,d\tau\right) ,
\end{equation}
where dispersion is induced via the convolution of the memory kernel with the time-delayed field amplitude (we omitted the quickly decaying initial value term here).

\section{Calculation of derivative of $r^+$}
\label{appendix:proof}

To derive \eqref{eq:lnrplus} we have to calculate
\begin{equation}
\partial_\omega\ln(r^+)=\frac{1}{r^+}\partial_\omega r^+
=\frac{1}{\Phi^-}\partial_\omega \Phi^--\frac{1}{\Phi^+}\partial_\omega \Phi^+,
\end{equation}
where we abbreviated $\partial/\partial \omega\equiv\partial_\omega$.
The coupled--wave equations \eqref{eq:Phi} together with their derivatives with respect to frequency
can be written as 
\begin{equation}
\begin{aligned}
\frac{\partial}{\partial z} \Phi^+(z) & = -i\hat\beta(\Omega)\Phi^+(z) -i\kappa\Phi^-(z),
\\
-\frac{\partial}{\partial z} \Phi^-(z) & = -i\hat\beta(\Omega)\Phi^-(z) -i\kappa\Phi^+(z),
\\
\frac{\partial}{\partial z} \partial_\omega\Phi^+(z) & =
-i\hat\beta(\Omega)\partial_\omega\Phi^+(z) 
-i\kappa\partial_\omega\Phi^-(z) -i\partial_\omega\beta(\Omega)\Phi^+(z) ,
\\
-\frac{\partial}{\partial z} \partial_\omega\Phi^-(z) & =
-i\hat\beta(\Omega)\partial_\omega\Phi^-(z) 
-i\kappa\partial_\omega\Phi^+(z)  -i\partial_\omega\beta(\Omega)\Phi^-(z) ,
\end{aligned}
\end{equation}
where $\hat\beta(\Omega) = \Delta\overline{\beta}(z)+\beta_{\mathcal{D}}(\Omega) +n_\textrm{g}\Omega/c$
and $\partial_\omega\beta(\Omega)$ is defined in \eqref{eq:dbetadeomega}.
By multiplying the left-- and right--hand sides of these equations by 
$\partial_\omega\Phi^-$, $\partial_\omega\Phi^+$, $-\Phi^-$, and  $-\Phi^+$, respectively, 
we get 
\begin{equation}
\begin{aligned}
\partial_\omega\Phi^-\,\frac{\partial}{\partial z} \Phi^+ & = 
-i\hat\beta(\Omega)\Phi^+\,\partial_\omega\Phi^-
-i\kappa\Phi^-\,\partial_\omega\Phi^-,
\\
-\partial_\omega\Phi^+\,\frac{\partial}{\partial z} \Phi^- & = 
-i\hat\beta(\Omega)\Phi^-\,\partial_\omega\Phi^+ 
-i\kappa\Phi^+\,\partial_\omega\Phi^+ ,
\\
-\Phi^-\,\frac{\partial}{\partial z} \partial_\omega\Phi^+ & =
i\hat\beta(\Omega)\Phi^-\,\partial_\omega\Phi^+ 
+i\kappa\Phi^-\,\partial_\omega\Phi^-
+i\partial_\omega\beta(\Omega)\Phi^-\,\Phi^+ ,
\\
\Phi^+\,\frac{\partial}{\partial z} \partial_\omega\Phi^- & =
i\hat\beta(\Omega)\Phi^+\,\partial_\omega\Phi^- 
+i\kappa\Phi^+\,\partial_\omega\Phi^+
+i\partial_\omega\beta(\Omega)\Phi^+\,\Phi^-.
\end{aligned}
\end{equation}
Adding all these equations implies
\begin{equation}
\frac{\partial}{\partial z}\left( 
\Phi^+\,\partial_\omega\Phi^-
-\Phi^-\,\partial_\omega\Phi^+ 
\right)
=
2i\partial_\omega\beta(\Omega)\Phi^+\,\Phi^-.
\end{equation}
An integration over the passive sections corresponding to the coordinate interval $[l,L]$ yields
\begin{equation}
\left[ 
\Phi^+\,\partial_\omega\Phi^-
-\Phi^-\,\partial_\omega\Phi^+ 
\right]\big|_{l}^{L}
=
2i\int_{l}^{L} \partial_\omega\beta(\Omega)\Phi^+\,\Phi^- dz. 
\end{equation}
Due to the reflecting  boundary condition, the expression 
$\left[\Phi^+\,\partial_\omega\Phi^--\Phi^-\,\partial_\omega\Phi^+ \right]$ 
vanishes at $z=L$ and 
\begin{equation}
-\Phi^+(l)\,\partial_\omega\Phi^-(l)
+\Phi^-(l)\,\partial_\omega\Phi^+(l) 
=2i\int_{l}^{L} \partial_\omega\beta(z,\Omega)\Phi^+(z)\Phi^-(z) dz
\end{equation}
follows.
Dividing by $-\Phi^+(l)\Phi^-(l)$  finally yields
\begin{equation}
\partial_\omega \log(r^+)
=-2i\frac{\int_{l}^{L} \partial_\omega\beta(z,\Omega)\Phi^+(z)\Phi^-(z)
  dz}{\Phi^+(l)\Phi^-(l)}.
\end{equation}

\section{The relation to static frequency chirp}
\label{appendix:chirp_reduction}

Here we justify the naming of the factor $F$ defined in \eqref{eq:F}.
We start from the roundtrip condition
\begin{equation}
\label{eq:roundtrip}
r^+(\tilde{\omega},g)r^-(\tilde{\omega},g)=1
\end{equation}
following from \eqref{eq:rpm} which can be transformed into equations for the modulus 
\begin{equation}
\Re{\left(\ln(r^-(\tilde{\omega},g))\right)}+\Re{\left(\ln(r^+(\tilde{\omega}))\right)}=0
\label{eq:roundtrip_real}
\end{equation}
and the phase
\begin{equation}
\label{eq:roundtrip_phase}
h(\tilde{\omega},g)\equiv\mathrm{Im}\big(\ln(r^-(\tilde{\omega},g))\big)+\mathrm{Im}\big(\ln(r^+(\tilde{\omega}))\big)
=2\pi m
\end{equation}
with $\tilde{\omega}=\omega-\omega_0$ and $m\in\mathbb{Z}$.
As a result of the solution of \eqref{eq:roundtrip} or, equivalently, \eqref{eq:roundtrip_real} and \eqref{eq:roundtrip_phase}, we obtain the real--valued frequency $\Omega=\tilde{\omega}$ and the gain $g$ of the steady lasing state.

Let the modal index in the active section vary due to a fluctuation of some parameter such as carrier density or temperature.
Eq. \eqref{eq:roundtrip_phase} establishes a relation between the index fluctuation $\delta \Delta n$ and the frequency fluctuation $\delta\omega$,
\begin{equation}
\delta h = \frac{\partial h}{\partial \omega}\delta\omega 
+ \frac{\partial h}{\partial n}\delta n= 0,
\end{equation}
where $\delta n \equiv \delta \Delta n$ and $1/\partial n \equiv 1/\partial \Delta n$.
The $\omega$--derivative of $h$ is
\begin{equation}
\frac{\partial h}{\partial \omega}
=\Im{\left(\frac{\partial\ln(r^-)}{\partial\omega}\right)}
+\Im{\left(\frac{\partial\ln(r^+)}{\partial\omega}\right)}
+\Im{\left(\frac{\partial\ln(r^-)}{\partial g}\right)}\frac{\partial g}{\partial\omega}.
\end{equation}
The $\omega$--derivative of the lasing gain $g$ can be determined from \eqref{eq:roundtrip_real},
\begin{equation}
\frac{\partial g}{\partial\omega}=-\frac{\Re{\left(\frac{\partial\ln(r^-)}{\partial\omega}\right)}+\Re{\left(\frac{\partial\ln(r^+)}{\partial\omega}\right)}}
{\Re{\left(\frac{\partial\ln(r^-)}{\partial g}\right)}}
=-\frac{1}{l}\Re{\left(\frac{\partial\ln(r^+)}{\partial\omega}\right)}
\label{eq:dgdo}
\end{equation}
taking into account
\begin{equation}
\frac{\partial\ln(r^-)}{\partial\omega}=-2il\frac{n_\mathrm{g}}{c}
\end{equation}
and
\begin{equation}
\frac{\partial\ln(r^-)}{\partial g}=l(1+i\alpha_\mathrm{H}),
\end{equation}
following from \eqref{eq:rpm} and \eqref{eq:betaoverline}, such that
\begin{equation}
\frac{\partial h}{\partial \omega}
=-2l\frac{n_\mathrm{g}}{c}+\Im{\left(\frac{\partial\ln(r^+)}{\partial\omega}\right)}
-\alpha_\mathrm{H}\Re{\left(\frac{\partial\ln(r^+)}{\partial\omega}\right)}
\end{equation}
is obtained.
Furthermore, \eqref{eq:rpm} and \eqref{eq:betaoverline} imply
\begin{equation}
\frac{\partial\ln(r^-)}{\partial n}=-2il\frac{2\pi}{\lambda_0}
\end{equation}
and
\begin{equation}
\frac{\partial h}{\partial n}=-\frac{4\pi l}{\lambda_0}.
\end{equation}
Therefore,
\begin{equation}
\delta h = - 2l\frac{n_\mathrm{g}}{c}\left[1-\frac{c}{2n_\mathrm{g}l}\Im{\left(\frac{\partial\ln(r^+)}{\partial\omega}\right)}+\alpha_\mathrm{H}\frac{c}{2n_\mathrm{g}l}\Re{\left(\frac{\partial\ln(r^+)}{\partial\omega}\right)}\right]\delta\omega - \frac{4\pi l}{\lambda_0}\delta n
= 0
\end{equation}
and
\begin{equation}
\delta\omega=-\omega_0\frac{\delta n}{n_\mathrm{g}}F^{-1}
\end{equation}
is gained.
For the solitary laser with $r^+(\tilde{\omega})$ taken at a fixed  $\tilde{\omega}=\Omega$, the same analysis results in
\begin{equation}
\label{eq:solitary}
\frac{\partial h}{\partial \omega}=-2l\frac{n_\mathrm{g}}{c}
\end{equation}
and
\begin{equation}
\delta\omega_\mathrm{s}=-\omega_0\frac{\delta n}{n_\mathrm{g}}.
\end{equation}
Therefore in a laser with a passive section or an external cavity, fluctuations of the frequency $\delta\omega$ due to a fluctuation of some parameter in the active section are reduced (or enhanced) by the factor 
\begin{equation}
\frac{\delta\omega}{\delta\omega_\mathrm{s}}=\frac{1}{F}
\end{equation}
compared to a the
solitary laser, as stated in, \textit{e.g.}, \cite{kazarinov1987relation, tromborg1987transmission}.
If the active section is not a single FP cavity, the chirp reduction factor can be generalized to \cite{tronciu2017instabilities}
\begin{equation}
F=1+\Re{\left((1-i\alpha_\mathrm{H})\frac{\frac{\partial\mathrm{ln}(r^+)}{\partial\omega}}{\frac{\partial\mathrm{ln}(r^-)}{\partial\omega}}\right)}.
\end{equation}


%

\end{document}